\def\fileversion{v2.25a} \def\filedate{2009/10/10}
\documentclass[edeposit,fullpage]{uiucthesis2009}
\usepackage{amsmath}
\usepackage[pdftex]{graphicx}
\usepackage{epstopdf}

\begin{document}

\title{Universal Sound Attenuation in Amorphous Solids at Low-Temperatures}
\author{Dervis Vural}
\department{Physics}
\schools{B.S, Middle East Technical University, 2002\\
         M.Sc, University of Illinois at Urbana-Champaign, 2004}
\phdthesis
\advisor{Anthony J. Leggett}
\degreeyear{2011}
\committee{Associate Professor Karin Dahmen, Chair\\Professor Anthony J. Leggett, Director of Research\\Professor David Ceperley\\Assistant Professor Aleksei Aksimentiev}
\maketitle

\frontmatter

\begin{abstract}
Disordered solids are known to exhibit quantitative universalities at low temperatures, the most striking of which is the ultrasonic attenuation coefficient $Q^{-1}(\omega)$. The established theory of tunneling two state systems (TTLS) in its original form (i.e. without extra fitting functions and parameters), is unable to explain this universality. While the TTLS model can be modified, particularly by including long range phonon induced interactions to explain the universal value of $Q^{-1}$, (a) it is not clear that the essential features of the original model that has been successful in explaining the experimental data is preserved, and (b) even if it is, it is not clear that the postulates of the original model remain \emph{necessary}.

The purpose of this study is to derive the universal acoustic absorption and related quantities observed in disordered solids by starting from a many-body quantum theory of \emph{unspecified} amorphous blocks that mutually interact through the strain field.

Based on very generic assumptions and having \emph{no adjustable fitting parameters}, the frequency and initial state averaged macroscopic attenuation $\langle Q^{-1}\rangle$ of a group of interacting disordered blocks is calculated in the low temperature regime ($T\ll\omega$) by a novel ``trace method'', which then is iterated up-to experimental length scales through a real space renormalization group approach. Then using a heuristic second-order perturbation argument, the frequency dependence of $Q^{-1}(\omega)$ is found, and combined with the previous result to yield the observed universal values in the MHz-Ghz range of frequencies.

It is concluded that the TTLS postulates are not necessary in order to explain, at least the thermal conductivity, velocity shift and sound attenuation of disordered media in the low temperature regime.

\end{abstract}

\begin{dedication}
A famous Zen koan asks, ``Two hands clap and there is a sound; what is the sound of one hand?''. The present work suggests, ``to logarithmic accuracy, it does not matter''.
\end{dedication}

\chapter*{Acknowledgments}

This project would not have been possible without the support of many people. Many thanks to my advisor, Anthony J. Leggett, who carefully and patiently read countless preliminary ideas which lead to the present one. Without his extraordinary physical intuition and formulation of the problem, this outcome would not have been possible. Thanks to my lovely wife P\i nar Zorlutuna. Having endured this process to its end with all the conversation that comes with it, she is now equally knowledgable about the field as I, if not more. Also, thanks to my mother and father for producing me, and persistently encouraging me to ``finish writing up''.

I acknowledge helpful discussions over many years with numerous colleagues, in particular Pragya Shukla, Karin Dahmen, Alexander Burin, Zvi Ovadyahu and Philip Stamp. Thanks to University of British Columbia, University of Waterloo, and Harvard University for generously providing accommodation during long visits.

Finally, thanks to all the coffee shops in Champaign, specifically Green Street Coffee Shop and Bar-Guilliani for generously nourishing me with espresso shots every day for free, for many years.
\\\\
This work was supported by the National Science Foundation under grant numbers NSF-DMR-03-50842 and NSF-DMR09-06921. 

\tableofcontents
\chapter{List of Symbols and Abbreviations}

\begin{symbollist}[0.7in]
\item[$l$] Phonon Mean-Free-Path
\item[$\lambda$] Wavelength
\item[$\chi_{ij:kl}$] Stress-Stress Linear Response Function
\item[$Q^{-1}$] Attenuation Coefficient ($\equiv l^{-1}\lambda/(2\pi^2)=(\pi\mbox{Re}\chi)^{-1})\mbox{Im}\chi$)
\item[$Q_0^{-1}$] Attenuation Coefficient of a microscopic block
\item[$\langle Q^{-1}\rangle$] Frequency and initial state averaged attenuation coefficient
\item[$r_0$] Size of a ``microscopic'' block
\item[$v_i$] Volume of Block i
\item[$v_{ij}$] Volume of Sphere with Radius Equal to Distance between Blocks i and j
\item[$T$] Temperature
\item[$T_0$] Reference Temperature
\item[$T_c$] Critical Temperature
\item[$\beta$] Inverse temperature
\item[$U_0$] Critical Energy / High-Energy Cutoff
\item[$U$] Renormalized $U_0$
\item[$N_b$] Number of Energy Levels Below Cutoff
\item[$\omega$] Probe Frequency
\item[$\omega_D$] Debye Frequency
\item[$\vec{k}$] Wavenumber
\item[$D(\omega)$] Phonon Density of States
\item[$f(\omega)$] Non-Phonon (block) Density of States
\item[$c_{\nu}(T)$] Phonon Specific Heat
\item[$C_{\nu}(T)$] Non-Phonon (block) Specific Heat
\item[$l,t$] Longitudinal, Transverse
\item[$c_l,c_t$] Longitudinal and Transverse Speed of Sound
\item[$K(T)$] Thermal Conductivity
\item[$\alpha$] Specific Heat Exponent
\item[$\beta$] Thermal Conductivity Exponent
\item[$\hat{H}$] Full Hamiltonian
\item[$\hat{H_{(ph)}}$] Phonon Hamiltonian
\item[$\hat{V}$] Phonon-Induced Stress-Stress Coupling
\item[$\hat{H_0^{s}}$] Noninteracting Block Hamiltonian of block $s$
\item[$\tau$] Pulse Duration
\item[$r_i$] Position of Block $i$
\item[$\hat{n}$] Unit Vector Along two Blocks
\item[$u$] Atomic Displacement
\item[$x_i$] Position Coordinate ($i=1,2,3$)
\item[$\rho$] Mass Density
\item[$e_{ij}$] Strain Tensor
\item[$\hat{T}_{ij}$] Stress Tensor Operator
\item[$L$] Block Linear Dimension
\item[$p_m$] Probability of Occupation of the $m^{th}$ state
\item[$Z$] Partition Function
\item[$z$] Renormalization step
\item[$\Lambda_{ijkl}$] Amorphous block Stress-Stress Coupling Coefficient
\item[p.v.] Principal Value
\item[TTLS] Tunneling Two Level System
\item[$\Omega$, $V$, $d$, $M_0$] TTLS Ground Frequency, Barrier Height, Well Separation, Mass
\item[$\Delta$, $\Delta_0$, $\gamma$] TTLS Well Asymmetry, Transition Matrix Element, Phonon Coupling Constant
\item[$I$, $I_0$] Sound Intensity, Critical Sound Intensity
\item[$P(\Delta,\Delta_0,\gamma)$] Probability distribution for TTLS parameters
\end{symbollist}

\mainmatter

\chapter{Introduction}

The generic term ``disordered solid'' (or ``amorphous solid'') is typically used to describe a large variety of condensed matter including metallic and dielectric glasses, disordered crystals, polymers, quasi-crystals and proteins. Starting from the pioneering work of Zeller and Pohl \cite{pohl71} it has become clear that the thermal, acoustic and electromagnetic response of disordered solids are (1) remarkably different than crystals, and (2) are remarkably similar among themselves. To put the problem posed by these two statements in proper context, it will be necessary to give a little background on the low temperature properties of both disordered and non-disordered condensed matter.

\section{A Comparison of Crystalline and Disordered Solids}
According to Debye's theory of crystalline insulators, at sufficiently low temperatures the amplitude $\vec{r}$ of atomic motion remains small enough to consider only the first order term in the inter-atomic force $F_i=2\mu_{ij} r_j$. Thus, the low temperature acoustic and thermal properties of all crystalline insulators are determined by a Hamiltonian which can be diagonalized into phonon modes, quantized with uniform increments of $E=\omega$.

In this framework, the standard text-book procedure \cite{kittel} to obtain crystalline specific heat is to simply integrate over these modes to find the average energy density,
\begin{eqnarray}\label{debye}
\langle E\rangle=\sum_p\int_0^{\omega_D} \omega D(\omega)n(\omega,\beta)d\omega,
\end{eqnarray}
where the summation is over phonon polarizations, and $n$ is the number of phonons at a given temperature $T=1/\beta$,
\begin{eqnarray}
n=\frac{1}{e^{\beta\omega}-1}
\end{eqnarray}
and $D$ is the phonon density of states per volume $v$, at wave number $\vec{k}$
\begin{eqnarray}
D(\omega)=\frac{|k|^2}{2\pi^2}\frac{d|k|}{d\omega}.
\end{eqnarray}
The upper cut-off $\omega_D$ of the integral in eqn(\ref{debye}) is the energy of a phonon with wavelength $\lambda$ equal to inter-atomic distance $a_0$. Such high end cutoffs are commonly used in condensed matter theories, including the present study (cf. Chapter 2), since they conveniently exclude smaller length scales for which the physical assumptions no longer hold. This of course, limits the validity of the theory to energy and temperatures below $\sim hc/a_0$.

In the continuum limit, $\lambda\gg a_0$, the linear dispersion relation 
\begin{eqnarray}
\omega_{l,t}=kc_{l,t}  
\end{eqnarray}
holds, where $c_l,t$ is the transverse $t$ or longitudinal $l$ speed of sound in the limit $\omega_{l,t}\to0$. At low temperatures $T\ll\omega_D$, eqn(\ref{debye}) can be easily integrated to yield the famous $T^3$ temperature dependence for the specific heat.
\begin{align}\label{famous}
c_v=\frac{\partial\langle E\rangle}{\partial T}=\frac{12\pi^4}{5}N\left(\frac{T}{\omega_D}\right)^3.
\end{align}

Let us now turn to the thermal conductivity, $K$. Since phonons are the main heat carriers of heat in a dielectric crystal, $K(T)$ can be calculated analogously to the kinetic theory of gases. If $\bar{l}$ and $\bar{c}$ is the frequency averaged mean free path and velocity of a phonon,
\begin{eqnarray}\label{K}
K\approx\frac{1}{3}c_v\bar{c}\bar{l}
\end{eqnarray}
Of course, for a plane wave propagating in a perfect infinite crystal whose atoms interact harmonically, $l$ and thus $K$ is infinite. However in practice, the phonons are restricted by the finite geometric scale $l=L$ of the sample. Thus, if there are no impurities and the temperature is low enough to neglect anharmonic interaction between atoms, from eqn(\ref{famous}) we roughly get $K\sim T^3$.

At first sight, the Debye theory seems equally applicable to disordered solids, since these materials are known to exhibit well-defined phonon modes, as seen from Brillouin scattering experiments below 33 GHz \cite{vacher76} and phonon interference experiments below 500 GHz \cite{rothenfusser83}. Thus, at energy scales for which the thermal or acoustic wavelength exceeds $a_0$ there is no a priori reason for the thermo-acoustic response of a solid, disordered or not, to deviate from that of the elastic continuum described above.

However, experimentally, disordered solids behave nothing like their crystalline counterparts (for two comprehensive reviews, see \cite{phillips81,esquinazi-rev}). The specific heat, measurements reveal\cite{pohl71} an super-linear temperature dependence of specific heat below $T_{0}$, 
\begin{eqnarray}
 C_v\propto T^{1+\alpha}
\end{eqnarray}
with $\alpha\sim0.1-0.3$. Remarkably, at $25mK$ this term dominates the Debye phonon contribution by a factor of about $1000$. Above $T_c$ 
the specific heat raises faster before reaching the Debye limit, which manifests as a bump in the $C_v/T^3$ vs $T$ curve (a historical convention used to compare the glassy specific heat to a ``crystalline reference''). $T_0$ is a more-or-less material dependent temperature, and typically varies between $1-30 K$.

According to specific heat measurements, it is clear that something other than phonons is being thermally excited in disordered solids. In the established model of amorphous solids, these are assumed to be ``tunneling two level systems''. In the present work, we acknowledge the presence of these degrees of freedoms but do not make any specific assumptions regarding their nature. A more complete description of both models will follow below.

Turning now to the thermal conductivity, measurements reveal a sub quadratic temperature dependence \cite{pohl71,anderson86}
\begin{eqnarray}
K\propto T^{2-\beta}.
\end{eqnarray}
with $\beta\sim0.05-0.2$. Again, around a temperature that empirically coincides with the ``bump'' in specific heat, $T_0$, this gradually changes to a temperature independent plateau. it is important to note in the present context, that even though $C_v$ and $K$ are traditionally fit to power laws, as we will see below, logarithmic corrections $C_v\sim T/\ln T$ and $K\sim T^2\ln T$, lead to indistinguishable functional forms below $T_0$)

The universality of disordered solids go as beyond these ``exponents'' in temperature dependences: In a disordered solid, heat is transported by phonons \cite{zaitlin75}, allowing the use of eqn(\ref{K}) to obtain the ultrasonic attenuation coefficient defined in terms of the phonon mean free path,
\begin{eqnarray}
 Q^{-1}\equiv \lambda/(2\pi^2l).
\end{eqnarray}
Below $T<T_0$, with $T_0$ again roughly coinciding with that defined above, the mean free path displays an amazing degree of universality\cite{anderson86,mb,pohl2002}: In this regime sound waves travel about 150 times their own wavelength, regardless of the chemistry or composition of the amorphous matrix. Above $T_0$, $\lambda/l$ rapidly increases to a (non-universal) constant of the order one.

Two other universalities are the ratios \cite{mb} of transverse and longitudinal sound velocity and phonon coupling,
\begin{eqnarray}
\frac{c_t}{c_l}\approx0.6\\
\frac{\chi_t}{\chi_l}\approx0.4
\end{eqnarray}

In the present work we will provide some theoretical justification to these ratios, and use their experimental values \cite{mb} as inputs to obtain the value of $Q^{-1}$, and related quantities.

In addition to these quantitative universalities, disordered solids display a series of very unusual acoustic non-linearities that are not present in crystals. For example, above a critical sound intensity $I_c$, the acoustic absorption is ``saturated''\cite{hunklinger72}; i.e. above $I_c$, the solid becomes transparent to sound:
\[\chi\sim\frac{1}{\sqrt{1+I/I_c}}.\]
A second example is the spectacular echo experiments. These are the acoustic analogs of the magnetic response of a spin particle in a nuclear magnetic resonance setup, where the amorphous sample is cooled down below $T<\omega$. Then two consequent pulses of equal frequency $\omega$ separated by time $\tau_s\gg\tau_p$ are applied for a duration of $\tau_p$ and $2\tau_p$ respectively. The two elastic pulses surprisingly induces a ``spontaneous'' third one, precisely $\tau_s$ after the second pulse.

It is remarkable that, although less pronounced, the universalities hold for metallic glasses as well. However, we limit our considerations in this thesis only to amorphous insulators, since the electron-electron and electron-phonon interactions add an extra layer of complexity to the problem. Further, we will focus only on the universality of \emph{linear} acoustic properties of disordered solids in the resonance regime, and the thermal conductivity below $T_c$. We will not discuss the nonlinear properties, or the acoustic response in the relaxation regime.

\section{A Brief Review of the Tunneling Two Level System Model}

The theoretical interpretation of the low temperature data on amorphous materials has for 40 years been dominated by the phenomenological ``tunneling two state system'' (TTLS) model \cite{ahv,p}, which has two main assumptions; the first is that, within all disordered solids there exists entities that tunnel between two metastable states $|L\rangle$ and $|R\rangle$ (conventionally called ``left'' and ``right'', although the tunneling coordinate is not specified; cf. below). The tunneling entity is described by a reduced double well Hamiltonian,
\[H_{TTLS}=\left(\begin{array}{cc}
   \Delta & -\Delta_0\\
-\Delta_0 & -\Delta
  \end{array}\right)+e\left(\begin{array}{cc}
   \gamma & 0\\
 0 & -\gamma
  \end{array}\right)\]
Here $\Delta$ is the double-well asymmetry, $\Delta_0$ is the tunneling matrix element expressed by the standard WKB formula
\begin{eqnarray}
\Delta_0\sim\Omega e^{-\xi}\nonumber\\
\xi=d\sqrt{2M_{0}V}
\end{eqnarray}
$M_{0}$ is the mass of the tunneling entity, $\Omega$ is the ground frequency of each (harmonic) well, $d$ is the separation between wells, and $V$ is the double well barier height. The TTLS couples to strain $e$ (and of course, the thermal phonons) through the coupling constant $\gamma$.

The second fundamental assumption of the TTLS theory is that the two level parameters of an ensemble of TTLS are distributed according to the probability density
\[P(\Delta,\Delta_0,\gamma)=\frac{\bar{P}\delta(\gamma-\gamma_0)}{\Delta_0}\]

The $1/\Delta_0$ factor can be obtained by assuming that the parameter $\xi$ in the exponent is uniformly distributed. $\gamma$ is assumed to be the same for all tunneling entities.

At the cost of introducing a fairly large number of fitting parameters, the TTLS theory gives an attractive explanation to the nonlinear acoustic effects\cite{phillips81} and a specific heat linear with temperature $c_v\sim T$ and quadratic in thermal conductivity $K\sim T^2$ \cite{ahv,p} reasonably close to experiment below $T_0$ (see however, Fig\ref{fig:Figure2} and corresponding discussion in Chapter 5). Moreover it gives the temperature and frequency dependences of the mean free path and sound velocity\cite{jackle72,golding76} (although the latter requires an additional fit function with no physical basis).

The dominant response of a TTLS to a sound wave in the regime $T<\omega$ occurs through ``resonant absorption''. A single phonon interacts with a TTLS with matching energy separation at thermal equilibrium. The mean free path is found from Fermi's golden rule for transition rates $1/\tau$. Using $l=c\tau$,
\begin{eqnarray}\label{mfp}
\lambda l^{-1}(T,\omega)\equiv2\pi^2Q^{-1}=2\pi^2\frac{\gamma^2\bar{P}}{\rho c^2}\mbox{tanh}\left(\frac{\omega}{T}\right)
\end{eqnarray}
this equation can be plugged in the Kramers-Kronig principal value integral
\begin{eqnarray}
c(T)-c(0)\equiv\Delta c=\frac{1}{\pi}\mbox{p.v.}\int_0^\infty\frac{c^2l^{-1}(\omega')}{\omega^2-\omega'^2}d\omega'
\end{eqnarray}
to yield a logarithmic velocity shift for $kT\sim\hbar\omega$ \cite{phyac76}
\begin{eqnarray}
\frac{\Delta c}{c}=-\frac{\gamma^2\bar{P}}{\rho c^2}\log\left(\frac{\omega}{T}\right) 
\end{eqnarray}
Since $T\ll\omega$ is experimentally difficult to attain, this equation is usually tested in the regime $T>\omega$ with respect to an (arbitrary) reference temperature $T_0$,
\begin{eqnarray}\label{tlsvel}
\frac{c(T)-c(T_0)}{c(T)}=\frac{\gamma^2\bar{P}}{\rho c^2}\log\left(\frac{T}{T_0}\right)
\end{eqnarray}
Therefore, (with one exception \cite{golding76}) most of our direct knowledge regarding $Q^{-1}$ (and thus its universality) in the resonance regime comes indirectly, through measuring the velocity shift at high temperatures and then obtaining $Q^{-1}$ from the Kramers-Kronig integral that goes the other way around.

For higher temperatures $T>\omega$, the dominant mechanism for sound absorption is quite different than that described above, through a ``relaxation'' process. This is when a low frequency sound wave modulates the TTLS energy $E$, causing the population ratio $n$ of excited to unexcited TTLS to differ from its equilibrium value
\[\bar{n}\propto e^{-\beta E}.\]
It is not difficult to see how acoustic energy is converted to heat this way: The pulse must do work to widen the levels of an excited TTLS, which then is ``lost'' when the TTLS emits a thermal phonon to decrease the overpopulated $n$. In the limit $\omega\tau_{min}\ll 1$ (which corresponds to higher temperatures), the theory predicts half the value of the absorption in the resonance regime, again independent of temperature and frequency
\begin{eqnarray}
Q^{-1}=\frac{\bar{P}\gamma^2}{2\rho c^2}
\end{eqnarray}
where $\tau_{min}$ is the minimum relaxation time among the TTLS with splitting $E=kT$.

\section{Problems with the Standard TTLS Model}
The first difficulty of the TTLS model is the uncertainty regarding its microscopic nature. Although the configuration coordinate is typically referred to as ``left'' and ``right'', neither the entity undergoing the tunneling motion, nor the nature of its motion is known. The TTLS may correspond to a single atom translating between two relatively low density regions in the solid, or two atoms may be sliding or rotating by each other. A recently discovered isotopic effect\cite{nagel04} proposes that the TTLS is as large as ten or twenty atoms.

Various mechanisms that could give rise to tunneling two levels were suggested for individual amorphous solids, such as the mixed phase metallic NbZr alloy domains\cite{lou76} or the hydroxyl ion impurities in vitreous silica \cite{phillips81a}. The tunneling impurities are also known to be present in disordered crystals, such as KBr-KCN solutions \cite{deyoreo86} and are extensively studied. However the following questions remain: First of all, even if two level excitations are present in (at least some) disordered media, obviously the are not the \emph{only} kind of excitation that is possible. Why then should the thermo-acoustic response of disordered media be entirely governed by \emph{two level} excitations? And furthermore, even if these excitations are present and dominant among all, it is not clear why their distribution of parameters $P(\Delta,\Delta_0,\gamma)$ should be so similar in every amorphous material.

Ideally, a theory for disordered condensed matter should start from assumptions that holds equally true for a very wide class of materials. Even if the TTLS hypothesis is one that is effectively correct, it seems important to base its assumptions on a firm physical basis.

Secondly, although the standard TTLS model can explain qualitative features such as phonon echoes and saturated absorption as well as approximate temperature dependences of quantities such as the specific heat and ultrasonic absorption \cite{phillips81,esquinazi-rev}, in its simplest form (i.e. without ad-hoc additional fit parameters and fit functions) it fails to explain velocity shift and absorption data at low temperatures\cite{classen00,fefferman08,golding76}, in addition to the the ``bump'' and ``plateau'' that appears in the specific heat $C_v(T)$ and thermal conductivity $K(T)$ data at temperatures above $T_c$\cite{pohl71,anderson86}. 

Thirdly, in the temperature regime where the TTLS model is supposed to work ($T<1K$), it fails to explain the quantitative and qualitative universalities \cite{leggett91,leggett88} the most striking one of which is the ultrasonic attenuation $Q^{-1}$ \cite{pohl2002}; it is not at all obvious\cite{anderson86,leggett88,leggett91} that the coefficient of eqn(\ref{mfp})
\begin{eqnarray}
\alpha=\frac{\gamma^2\bar{P}}{\rho c^2}
\end{eqnarray}
should be material independent. In fact, the quantities in the numerator and denominator are independent parameters of the TTLS theory, and if we restrict ourselves to dielectric glasses and polymers, while $\mu_l=\rho c_l^2$ varies from material to material by nearly three orders of magnitude, $Q^{-1}$ remains around $(3\pm2)\times10^{-4}$. It seems extremely unlikely that $\gamma^2$ and $\bar{P}$ happen to correlate with $\mu$ to this extent by pure coincidence.

Finally, the model, in its original form, neglects the fact that, as a result of interaction with the the phonon field, the stress of each TTLS's must be coupled\cite{joffrin75}. Although the original paper was formulated in the language of the TTLS, its outcome is neither sensitive to the number of levels of the interacting excitations, nor to the probability distribution function of their parameters. The only requirement is that the stress is coupled to the phonon field linearly (c.f. Appendix-A).

Recently this interaction was incorporated in the TTLS paradigm\cite{burin96}, and was shown that it leads to the experimentally observed small universal value of $Q^{-1}$. However, it seems legitimate to ask whether the original features of TTLS, such as those that give rise to saturation of absorption and echoes will survive this significant modification. Furthermore, it may be that the original assumptions of the TTLS model may be \emph{unnecessary} after this modification. It is possible that the presence of elastic interactions gives rise to the anomalous glassy behavior, regardless of what is interacting.

In \cite{leggett88,leggett91} it was conjectured that if one starts from a \emph{generic} model in which at short length scales there is a contribution to the stress tensor from some anharmonic degrees of freedom, and take into account their phonon mediated mutual interaction, one will obtain the significant features of glasses below 1K. The goal of this project is to quantitatively justify this conjecture, by calculating the frequency and temperature dependence of $Q^{-1}$ and the quantities related to it in the low temperature ``resonance'' regime $T\ll\omega$. The relaxation regime, non-linear effects or the intermediate temperature ($T>T_c$) behavior will not be considered in this work.

The layout of the thesis is as follows: In chapter-2 the precise details of the model is defined, and the central object of the study, namely, the dimensionless stress-stress correlation, whose thermally-averaged imaginary part is the measured ultrasonic absorption $Q^{-1}$ is introduced. In section 3 a real-space renormalization calculation of the \emph{average} of $Q_m^{-1}(\omega)$ over the frequency $\omega$ and the starting state $m$ (for details of the notation see below) is carried out. Here it is shown that this quantity vanishes logarithmically with the volume of the system and for experimentally realistic volumes, and has a small value $\sim0.015$. In section 4, on the basis of a heuristic calculation up to second order in the phonon-induced interaction, it is argued that the functional form of $Q^{-1}(\omega)$ at $T=0$ should be ($\mbox{ln}\omega)^{-1}$, and that when we combine this result with that of section 3, the numerical value of $Q^{-1}$ for experimentally relevant frequencies should be universal up to logarithmic accuracy and numerically close to the observed value $3\times10^{-4}$. In section 5 we attempt to assess the significance of our calculations.

\chapter{The Model}
Imagine the disordered solid as being composed of many (statistically identical) cubes of size $L$. The precise value of $L$ need not be specified, so long as it is much smaller than the experimental sample size, yet much larger than the (average) inter-atomic distance $a$. For this system we can define the strain tensor as usual
\begin{eqnarray}\label{2.1.1}
 e_{ij}=\frac{1}{2}\left(\frac{\partial u_{i}}{\partial x_j}+\frac{\partial u_j}{x_i}\right)
\end{eqnarray}
where $\vec{u}(\vec{r})$ denotes the displacement relative to some arbitrary reference frame of the matter at point $\vec{r}=(x_1,x_2,x_3)$. 

The Hamiltonian can be defined as a Taylor expansion up-to terms first order in strain $e_{ij}$
\begin{eqnarray}\label{2.1.2}
\hat{H}=\hat{H}_0+\sum_{ij}e_{ij}\hat{T}_{ij}+\mathcal{O}(e^2)
\end{eqnarray}
where the stress tensor operator $\hat{T}_{ij}$ is defined by
\begin{eqnarray}\label{2.1.3}
\hat{T}_{ij}=\partial \hat{H}/\partial e_{ij}
\end{eqnarray}
Note that, in general, in a representation in which $\hat{H}_0$ is diagonal, $\hat{T}_{ij}$ will have both diagonal and off-diagonal elements.

We can define the static elasticity modulus $\chi^{(0)}$, a fourth order tensor, as
\begin{eqnarray}\label{2.1.4}
\chi^{(0)}_{ij:kl}\equiv \frac{1}{L^3}\left.\frac{\partial\langle\hat{T}_{ij}\rangle}{\partial e_{ij}}\right|_{e(T)}\equiv \frac{1}{L^3}\left.\left\langle\frac{\partial^2\hat{H}}{\partial e_{ij}\partial e_{kl}}\right\rangle\right|_{e(T)}
\end{eqnarray}
where the derivative must be taken at the thermal equilibrium configuration. For $L\gg a$ an amorphous solid is rotationally invariant. Due to this symmetry, any component of $\chi_{ij:kl}^{(0)}$ can be written in terms of two independent constants \cite{landau-elastic}, for which we pick the transverse $\chi_t$ and longitudinal $\chi_l$ response;
\begin{eqnarray}\label{2.1.5}
\chi^{(0)}_{ij:kl}=(\chi_l-2\chi_t)\delta_{ij}\delta_{kl}+\chi_t(\delta_{ik}\delta_{jl}+\delta_{il}\delta_{jk})
\end{eqnarray} 
In the approximation of an elastic continuum, these are related to the velocities $c_l$ and $c_t$ of the corresponding longitudinal and transverse sound waves (of wavelength $\lambda$ such that $a\ll\lambda\ll L$) by
\begin{eqnarray}\label{2.1.6}
\chi_{l,t}=\rho c_{l,t}^2
\end{eqnarray}
where $\rho$ is the mass density of the material.  As expected, the  continuum approximation leads to $Q^{-1}=0$. Clearly, we must take into account a Hamiltonian more general than (\ref{2.1.2}) to describe the acoustic properties of disordered solids.

\section{The Stress-Stress Correlation Function}

To go beyond the continuum approximation, we add an arbitrary ``non-phonon'' term $\hat{H}'(e_{ij})$ to eqn(\ref{2.1.2}),
\begin{eqnarray}\label{2.2.3}
 \hat{H}(e_{ij})\equiv \hat{H}_{el}(e_{ij})+\hat{H}'(e_{ij})
\end{eqnarray}
Here, the purely elastic contribution $H_{el}$, is given by
\begin{eqnarray}\label{2.2.1}
 \hat{H}_{el}(e_{ij})=\mbox{const.}+\int \frac{1}{2}d^3r\sum_{ijkl}\chi_{ij:kl}^{(0)}e_{ij}(\vec{r})e_{kl}(\vec{r})+\frac{1}{2}\sum_i\rho\dot{\vec{u}}_i^2(\vec{r})
\end{eqnarray}
Of course, the second term is meaningful if the velocity field $\dot{\vec{u}}$ is slowly varying over distances $a$. The ``non-phonon'' term $H'(e_{ij})$ is completely general; we neither assume that it is small compared to the purely elastic contribution $H_{el}$, nor do we identify it with any particular excitation or defect. As above, we define the ``elastic'' contribution to the stress tensor $\hat{T}_{ij}$ by
\begin{eqnarray}\label{2.2.2}
 \hat{T}_{ij}^{(el)}\equiv\sum_{ijkl}\chi_{ij:kl}^{(0)}e_{kl},
\end{eqnarray}
and as we have done in (\ref{2.1.2}) and (\ref{2.1.3}) we can define the ``non-phonon'' contribution to the stress tensor by 
\begin{eqnarray}\label{2.2.4}
 \hat{H}'&=&\hat{H}'_0+\sum_{ij}e_{ij}\hat{T}_{ij}'+\mathcal{O}(e^2)\\
\hat{T}_{ij}'&=&\partial \hat{H}'/\partial e_{ij}
\end{eqnarray}
Note that the strain $e_{ij}$ may be due to thermal phonons, as well as experimental probing strains. Since all the contribution to $Q^{-1}$ comes from the non-phonon contributions, for the sake of simplifying our notation we will omit the primes in $\hat{H}_0'$ and $\hat{T}_{ij}'$ from now on, and used the unprimed symbols $\hat{H}_0$ and $\hat{T}_{ij}$ to denote non-phonon contributions.

We will now define the non-phonon linear response function at scale $L$. A sinusoidal strain field with infinitesimal (real) amplitude $e_{ij}$, 
\begin{eqnarray}\label{2.2.6}
e_{ij}(\vec{r},t)=e_{ij}[e^{i(\vec{k}.\vec{r}-\omega t)}+e^{-i(\vec{k}.\vec{r}-\omega t)}]
\end{eqnarray}
will give rise to a stress response, $\langle T_{ij}\rangle$ (which in general is complex):
\begin{eqnarray}\label{2.2.7}
\langle T_{ij}\rangle(\vec{r},t)=\langle T_{ij}\rangle e^{i(\vec{k}.\vec{r}-\omega t)}+\langle T_{ij}\rangle^* e^{-i(\vec{k}.\vec{r}-\omega t)},
\end{eqnarray}
The complex linear response function $\chi_{ij:kl}(\vec{q},\omega)$ is defined in the standard way
\begin{eqnarray}\label{2.2.8}
 \chi_{ij,kl}(q,\omega)=\frac{1}{V}\frac{\partial \langle T_{ij}\rangle(\vec{q},\omega)}{\partial e_{kl}}
\end{eqnarray}
In practice, we will be interested in the $\lambda\gg a$ limit. We may therefore work in the linear dispersion limit
\begin{eqnarray}
\chi(\omega,\vec{q})\approx\chi(\omega,\omega/c_{l,t})\equiv\chi(\omega) 
\end{eqnarray}

It is not immediately obvious that the non-phonon response function $\chi_{ij:kl}(\vec{q},\omega)$ will have the isotropic form analogous to (\ref{2.1.5}), especially at the latter stages of the renormalization where we will be considering amorphous cubes with sizes comparable to the wavelength; however, since it is clear that any complications associated with this consideration are sensitive at our arbitrary choice of building-block shape, we will assume that a more rigorous (q-space) calculation will get rid of them, and thus assume that $\chi_{ijkl}(\omega)$ will have the same isotropic form as (\ref{2.1.5}), thereby defining ``longitudinal'' and ``transverse'' response functions $\chi_{l,t}(\omega)$ for cubes of size $L\gg a$.

We may now calculate the absorption time $\tau^{-1}=cl^{-1}$, in terms of the imaginary part of the linear response function. From Fermi's Golden rule,
\begin{eqnarray}\label{2.2.9}
 Q_\alpha^{-1}(\omega)\equiv\frac{\lambda}{2\pi^2l} =\frac{1}{\pi\rho c_\alpha^2}\mbox{Im}\chi_\alpha(\omega)
\end{eqnarray}
where the quantity $\mbox{Im}\chi_\alpha(\omega)$ is given explicitly, in the representation in which $\hat{H}_0$ is diagonal, by the formula
\begin{eqnarray}
\mbox{Im}\chi_{ij:kl}(\omega)=\sum_mp_m\chi^{(m)}_{ij:kl}(\omega)\label{2.2.10}\\
\chi_{ij:kl}^{(m)}(\omega)=\frac{\pi}{L^3}\sum_n\langle m|T_{ij}|n\rangle\langle n|T_{kl}|m\rangle\delta(E_{n}-E_{m}-\omega)\label{2.2.11}
\end{eqnarray}
where $|m\rangle$ and $|n\rangle$ denote exact many-body eigenstates of $\hat{H}_{0}$, with energies $E_m$, $E_n$, and $p_m$ is the probability that at thermal equilibrium the system initially occupies state $m$,
\begin{eqnarray}
p_m=\frac{1}{Z_\beta}e^{-\beta E_m}
\end{eqnarray}
where $Z$ is the partition function.

Our main objective is to calculate (\ref{2.2.10}), which depends on many-body energy levels and stress tensor matrix elements. Note that the formula is quite general; substituting TTLS assumptions and parameters in it directly gives eqn(\ref{mfp}). 

As an interesting side note, we point out that the TTLS mean free path is independent of frequency at zero temperature, which causes the Kramers-Kronig integral
\begin{align}\label{kkdiscuss}
\frac{\Delta c}{c}=\frac{2}{\pi}\int_0^\infty\frac{\mbox{Im}\chi(\omega')}{\omega'}d\omega'
\end{align}
to diverge logarithmically at zero frequency. This suggests that the actual zero temperature form of $Q_{\alpha}(\omega)$ is a weakly decreasing function of decreasing $\omega$. We will discuss this matter further in Chapter-4, and propose that the actual frequency dependence is $\sim\log^{-1} (U/\omega)$, which is the closest form to a constant that prevents (\ref{kkdiscuss}) from diverging\footnote{strictly speaking, we require $Q^{-1}(\omega)=\lim_{\epsilon\to0^+}1/\log^{1+\epsilon}U/\omega$.}

\section{Virtual Phonon Exchange between Blocks}
Let us define a ``block'' to be all the non-phonon degrees of freedom in a region enclosed by a cube of volume $L^3$ and consider a large collection of \emph{bare} uncorrelated blocks as described by the Hamiltonian (\ref{2.2.4}). Since the strain $e_{ij}$ includes the phonon field, the exchange of phonons must give rise to an effective coupling between pairs of block stress tensors. The phonon degrees of freedom are harmonic; therefore the stress-stress coupling should have the generic form
\begin{eqnarray}\label{2.3.1}
 H_{int}^{(12)}=\int_{V_1}d\vec{r}\int_{V_2}d\vec{r}'\sum_{ijkl}\Lambda_{ijkl}(\vec{r}-\vec{r}')T_{ij}(\vec{r})T_{kl}(\vec{r'}).
\end{eqnarray}
The function $\Lambda_{ijkl}(\vec{r}-\vec{r}')$ is calculated in the paper of Joffrin and Levelut\cite{joffrin75}. For ``large'' $r-r'$. it has the form\cite{esquinazi-rev}
\begin{eqnarray}\label{interaction}
 \Lambda_{ijkl}(\vec{r}-\vec{r}')=\frac{1}{\rho c^2_t}\frac{1}{2\pi|\vec{r}-\vec{r}'|^3}\tilde{\Lambda}_{ijkl}(\vec{n})\label{2.3.2}\nonumber\\
\tilde{\Lambda}_{ijkl}=-(\delta_{jl}-n_jn_l)\delta_{ik}+\left(1-\frac{c_t^2}{c_l^2}\right)[-\delta_{ij}\delta_{kl}-\delta_{ik}\delta_{jl}-\delta_{il}\delta_{jk}\nonumber\\
+3(\delta_{ij}n_kn_l+\delta_{ik}n_jn_l+\delta_{il}n_kn_j+\delta_{jk}n_in_l+\delta_{jl}n_in_k+\delta_{kl}n_in_j)-15n_in_kn_kn_l]\label{2.3.3}
\end{eqnarray}
where $\vec{n}$ is the unit vector along $\vec{r}-\vec{r}'$. The interaction ceases to have the $1/r^3$ form for length scales smaller than $r_0$, where
\begin{align}\label{cutoff}
\langle V(r_0)\rangle\sim\frac{hc}{r_0}\equiv U_0.
\end{align}
Beneath this length-scale (beyond this energy scale) the interaction becomes oscillatory, and this can be taken into account by introducing a cutoff energy level $U_0$ to any integral or sum over energies. In the present theory, the cutoff energy associated with the ``microscopic'' blocks is one of the input parameters, and will be discussed further in the next section.

Due to condition (\ref{cutoff}), the virtual phonon wavelengths considered in any stage of this calculation will be larger than the block size. Thus, we will assume that $\vec{r}-\vec{r'}$ can be replaced with the distance between the center points of two blocks $R_1-R_2$, and that the stress \[T_{ij}\approx\int_{v_s}\hat{T}_{ij}(\vec{r})d\vec{r}\] is \emph{uniform} throughout a single block. Then, the interacting many-body Hamiltonian for a collection of blocks can be written as
\begin{eqnarray}\label{2.3.4}
 \hat{H}_N=\sum_{s=1}^N\hat{H}_0^{(s)}+\sum_{\substack{s,s'=1\\s< s'}}^N\sum_{ijkl}\Lambda_{ijkl}(\vec{R}_s-\vec{R}_s')T_{ij}^{(s)}T_{kl}^{(s')}.
\end{eqnarray}
Eqn(\ref{2.3.4}) represents the Hamiltonian $H_0$ of the ``super block'' (of side $\sim N^{1/3}L$) composed by the $N$ blocks of side $L$; in principle, we can redefine the energy levels and stress tensor matrix elements for a super block and iterate the procedure untill we reach experimentally realistic length scales. At a given stage of renormalization, every pair $T^{(s)}T^{(s')}$ comes together with a $\Lambda$ which is proportional to $L^{-3}$ (note the same factor in the definition of $\chi$). Thus it is not difficult to see that the procedure is scale invariant, and the renormalization group equation of $Q^{-1}$ might converge to a fixed point with increasing volume.

Given infinite computational resources, one could start from arbitrary microscopic forms of $\hat{H}_0$ and $\hat{T}_0$, turn-on the interactions (\ref{2.3.1}) between blocks, diagonalize the N-body Hamiltonian $H_N$, and $T_{ij}=\partial H_N/\partial e_{ij}^{(N)}$ and iterate the procedure to see whether or not the final forms $H_N$ and $T_{ij}$ at the macroscopic scale are independent of the starting $\hat{H}_0$ and $\hat{T}_0$.

Unfortunately, the number of levels and matrix elements grow exponentially with increasing volume. Therefore it is not feasible to solve the problem stated in this form. We will instead narrow down our input parameters and focus on few output observables that can be deduced analytically.

\section{Input Parameters}

We believe that it is one of the strengths of the present work that our results do not rely on adjustable parameters, or the existence of \emph{other} microscopic (unmeasurable) universal ratios to explain the observable one \cite{yu89, wolynes01,turlakov04,parshin94,stamp09} (though cf.\cite{burin96}). The only two inputs on which our outcome depends sensitively are the ratios $c_l/c_t$ and $\chi_l/\chi_t$ (cf. below for details of the notation) both of which are observed \emph{experimentally} to vary little between different amorphous systems (cf. also Appendix). Our third input $r_0$, which is the size of a ``microscopic amorphous block'' (defined below) only enters into our equations logarithmically.

\subsection{Microscopic Building-Block Size}
As discussed in the introduction, the temperature dependence of thermal conductivity and specific heat changes at a critical temperature $T_c\sim 1K-30K$. At $10K$ the dominant phonons have wavelengths of the order $50\AA$, and in \cite{leggett91} it is argued that this is just the scale at which we get a crossover from ``Ising'' to ``Heisenberg'' behavior. Essentially, it is this length scale at which the approximations used in obtaining the simple $R^{-3}$ form of $\Lambda$, eqn(\ref{2.3.1}) breaks down. Therefore, we take the microscopic starting size to be about $r_0\sim 50\AA$, which is still much greater than $a$. We identify the ultraviolet cutoff $U_0$ with $T_c$, and expect the universality to break down as the block size becomes comparable to atomic size. 

$U_0$ is the only quantity in the present work that is not \emph{directly} measurable. However we should emphasize that (a) strictly speaking $U_0$ is not a free parameter, since on a priori grounds we can assign an approximate value to it (b) it can be experimentally obtained \emph{indirectly}, from thermal conductivity and specific heat data through $T_c=hc/r_0$. Most importantly, (c) our results, if they depend at all, depend on $U_0$ only logarithmically.

\subsection{Meissner-Berret Ratios}
The second and third input parameters we will be using are the ratio of longitudinal to transverse speed of sound $c_t/c_l$ and phonon coupling constants $\chi_t/\chi_l$, which are known to be universal among materials up to a factor of 1.2 (cf. ref\cite{mb}, Fig.1 and 3). Even though we use as inputs the experimentally obtained values, it is not difficult to produce first-principle theoretical justifications for either (cf. below).

Suppose that the inter-atomic interactions (which reduce to (\ref{2.2.1}) for small strains) is due to some arbitrary inter-atomic (or inter-block) interaction $\phi(r)$. Namely,
\begin{eqnarray}
 \hat{H}_{el}(e_{ij})=\mbox{const.}+\sum_{ab}^N\phi(r_{ab})+\frac{1}{2}\sum_i\rho\dot{\vec{u}}_i^2(\vec{r})
\end{eqnarray}
According to the virial theorem, the potential energy expectation value of a harmonic degree of freedom is equal to that of the kinetic energy, therefore for the purposes of obtaining the ratio $c_l/c_t$ without loss of generality we will drop the latter, and the const. term.
\begin{eqnarray}
 \langle H\rangle=\sum_{ab}\langle\phi_{ab}\rangle.
\end{eqnarray}
Further, suppose that the relative displacement $u_{ab}$ of two blocks $a$ and $b$ are proportional to the distance $r_{ab}$ between them. By definition of $e$, it follows that.
\begin{eqnarray}\label{trivial}
 u_{ab,x}=e_{xy}r_{ab,y}\\
 u_{ab,x}=e_{xx}r_{ab,x}
\end{eqnarray}
While this is clearly true for length scales for which $r\gg a$, in an amorphous structure large deviations from this may occur locally for $r$ of the same order as $a$. However these should even out in the thermodynamic limit $N\to\infty$.

The speed of sound is related to the real part of the zero frequency response function according to (\ref{2.1.6}).
\begin{eqnarray}\label{mb-main}
\chi_{0t,l}\approx\left.\frac{\partial^2}{\partial e_{ij}^2}\sum_{a<b}^N\langle\phi(r_{ab})\rangle\right|_{e_{ij}=0}=\sum_{a<b}^N\left[\frac{\partial|r_{ab}|}{\partial e_{ij}^2}\frac{\partial \phi(r_{ab})}{\partial r_{ab}}+\frac{\partial^2\phi(r_{ab})}{\partial r_{ab}^2}\left(\frac{\partial r_{ab}}{\partial e_{ij}}\right)^2\right]_{e_{ij}=0}
\end{eqnarray}
where $i\neq j$ and $i=j$ give $\chi_t$ and $\chi_l$ respectively. The first term of the right hand side must be zero due to stability requirements. Thus all we need to do is to substitute (\ref{trivial}) into the second term. For a purely transverse strain
\begin{eqnarray}
 r_{ab}=\sqrt{(r_{ab,x}+e_{xy}r_{ab,y})^2+r_{ab,y}^2+r_z^2},
\end{eqnarray}
which can be differentiated twice and substituted in (\ref{mb-main}). Letting $\sum_{ab}\to L^{-3}\int r^2drd\Omega$,
\begin{eqnarray}\label{chit}
\chi_{0t}=\frac{1}{L^3}\int \frac{\phi''(r)}{r^2}r^2dr\int r_x^2r_y^2 d\Omega.
\end{eqnarray}
where $\Omega$ is the solid angle. Similarly, for a purely longitudinal strain
\begin{eqnarray}
r_{ab}=\sqrt{(r_{ab,x}+e_{xx}r_{ab,x})^2+r_{ab,y}^2+r_z^2}.
\end{eqnarray}
Doing the same as above, we get
\begin{eqnarray}\label{chit}
\chi_{0l}=\frac{1}{L^3}\int \frac{\phi''(r)}{r^2}r^2dr\int r_x^4 d\Omega.
\end{eqnarray}
Then, the ratio for the speeds of sounds only depend on the angular integrals,
\begin{eqnarray}
 \frac{c_t}{c_l}=\sqrt{\frac{\chi_{0t}}{\chi_{0l}}}=\frac{1}{\sqrt{3}}
\end{eqnarray}
which is $6\%$ larger than the experimental (average) value.

The second ratio $\mbox{Im}\chi_t/\mbox{Im}\chi_l$ is not as trivial to obtain, since these ratios come from the non-phonon degrees of freedom. However an argument similar to the above can be made if we consider a second order perturbation expansion of the ground state in the interaction $V$, and differentiate it twice with respect to strain $e$. Then the (short range) interaction $\phi(r)$ must be replaced with the square of the (long range) elastic coupling (\ref{2.3.1}), and the above argument goes through.

\subsection{Many-Body Density of States}
On general grounds, we can assume that the normalized density of states of the interacting and noninteracting system, can be written as a power series, 
\begin{eqnarray}
f(\omega)=\sum_kc_k\omega^k\label{dosassumption1}\\
f_0(\omega)=\sum_kc_{0k}\omega^k\label{dosassumption2}
\end{eqnarray}
with dominating powers much larger than unity. Note that the density of states $F(E)$ of a composite system is given by the convolution of the density of states $f_i(E)$ of the constituent objects;
\begin{eqnarray}
F(E)=(f1*f2*f3*\ldots)(E)
\end{eqnarray}
where
\begin{eqnarray}
(f1*f2)(E)\equiv\int_{-\infty}^{\infty}f(E-\omega')f(\omega')d\omega'
\end{eqnarray}
From which it follows the dominating power of the density of states must be proportional to the number of particles if the system is extensive in energy. Thus, (\ref{dosassumption1}) and (\ref{dosassumption2}) holds quite generally.

Finally, while the actual many-body density of states we deduce from the specific heat data (cf. Appendix-B), 
\begin{align}\label{2.4.11}
f(E)= \mbox{const.}e^{(NE/\epsilon_0)^{1/2}}
\end{align}
is consistent with the form assumed above, as we will see, none of our results will be sensitive to the precise choice of $c_n$.

\subsection{Initial State Dependence}
Finally, we will specify the initial state $|m\rangle$ dependence of the response function, which will be the simplest possible choice consistent with our general assumptions, namely the ``random form'',

\begin{eqnarray}\label{isi}
\chi_\alpha^{(m)}(\omega)=\mbox{const}.\theta(E_m+\omega)=(\rho c_\alpha^2)Q_0^{-1}\theta(E_m+\omega)
\end{eqnarray}

Important: This form will not be used until we consider the frequency dependence of $Q^{-1}$ in Chapter 4 (cf. eqn. \ref{freqdependence}). Any conclusion we reach in Chapter 3 is not sensitive to this assumption!

It should be carefully noted that the TTLS form of $\chi_\alpha^{(m)}$ is \emph{not} a special case of eqn(\ref{isi}); this may be seen by noting that the form of $Q^{-1}$ given by the latter is approximately,
\begin{eqnarray}\label{mfpdcv}
Q^{-1}(\omega,T)=Q_0^{-1}(1-e^{-\omega/T}).
\end{eqnarray}
which is different from eqn (\ref{mfp}), though not qualitatively so. An important point that should be emphasized is, if the non-phononic hamiltonian consisted entirely of harmonic oscillators $Q^{-1}$ would be independent of temperature, which is qualitatively very different than eqn(\ref{mfp}). Intuitively, the ansatz (\ref{mfpdcv}) describes a model intermediate between a harmonic-oscillator and the TTLS one, but in some sense close to the latter. In principle, with the knowledge of the many body density of states (and thus partition function), a more accurate functional form for the initial state dependence can be found. It is our hope that a more realistic m-dependence does not qualitatively alter the calculations that will follow in the next sections.
 
\chapter{The Universality of the Average Attenuation}
We introduce this chapter by discussing a simpler system (cf. \cite{fisch80,leggett91}) in which a collection of spin-like objects couple to strain through a coupling coefficient $\gamma$. As a result of this interaction we will get an effective ``spin-spin'' interaction which is roughly of the form $g/r^3$, where \[g=\frac{\eta \gamma^2}{\rho c^2}\]
with $\eta$ a dimensionless number of order 1. If the single-spin excitation spectrum $\bar{P}$ is assumed to be independent of energy, it follows on dimensional grounds \cite{fisch80} that \[\bar{P}\propto\frac{1}{g}.\]
In this model, the dimensionless attenuation coefficient is simply \[Q^{-1}=\frac{\pi\gamma^2\bar{P}}{2\rho c^2}.\]
Furthermore if we include all phonon modes (longitudinal and transverse) this number if reduced by a factor of 3. Our purpose in this chapter is to generalize this argument \cite{fisch80} to a more generic model. We will (a) not necessarily assume ``single particle'' excitations, and (b) take into account not only different phonon modes, but all tensor components, and show that (a) and (b) alone lead to a surprisingly small value of frequency and initial state averaged attenuation, $\langle Q^{-1}\rangle$.
\section{Coupling Two Generic Blocks}
The central quantity we will be interested in this sectionis the frequency and initial state averaged ultrasound attenuation coefficient, defined as
\[\langle Q_0^{-1}\rangle=\frac{1}{U_0N_b}\sum_{n}\int_0^{U_0} Q_{n}^{-1}(\omega-E_n)d\omega\]
where $N_b$ is the number of levels of the block, and $U_0$ is an energy level of the order $U_0=hc_\alpha/L$, but the precise value is not essential for our purposes below.

Our strategy will be to evaluate the quantity $M=\mbox{Tr}(V^2)$ in the $H_0$ and $H_0+V$ eigenbasis. Let us start by considering two blocks only, labeled by $1$ and $2$. We will follow Einstein's summation convention for tensor indices.
\begin{eqnarray*}
\hat{V}^2&=&\left(\Lambda_{ijkl}\hat{T}_{1,ij}\hat{T}_{2,kl}\right)^2\\
&=&\Lambda_{ijkl}\Lambda_{i'j'k'l'}T_{1,ij}T_{1,i'j'}T_{2,kl}T_{2,k'l'}
\end{eqnarray*}
If $H_0$ has eigenvectors $\{|n_0\rangle\}$ $M$ can be evaluated as
\begin{align}\label{tr2}
 M=\sum_{mn}\Lambda_{ijkl}\Lambda_{i'j'k'l'}\langle m_0|T_{1,ij}T_{1,i'j'}|n_0\rangle\langle n_0|T_{2,kl}T_{2,k'l'}|m_0\rangle
\end{align}
Remember that $\Lambda$ depends on the relative positions of block 1 and 2.

Any expression of the form
\begin{align}
I= \sum_nF(E_{n}-E_{m})\langle n_0 |T_{ij}|m_0 \rangle\langle n_0 |T_{kl}|m_0 \rangle
\end{align}
can be written in terms of an integral of $\chi_0$ by inserting unity inside the sum,
\begin{align}\label{unity}
I&=\int\sum_nF(E_{n}-E_{m})\delta(E_{n}-E_{m}-\omega)\langle n_0 |T_{ij}|m_0 \rangle\langle n_0 |T_{kl}|m_0 \rangle d \omega\\
&=\int F(\omega)\chi_{0m,ijkl}(\omega)d\omega
\end{align}
Doing this twice in eqn(\ref{tr2}) the the trace can be written in terms of $\chi_0(\omega)$
\begin{eqnarray}\label{trace}
M=v_1v_2\sum_{n_1n_2}\int\int\Lambda_{ijkl}\Lambda_{i'j'k'l'}\chi_{0,n_1,iji'j'}(\omega')\chi_{0,n_2,klk'l'}(\omega'')d\omega'd\omega''
\end{eqnarray}
Here $v_1$ and $v_2$ are the volumes of block $1$ and $2$ respectively and both the sums and integrals are over the whole spectrum. A brief reminder of notation: The first subscript 0 means $\chi$ is the response of ``noninteracting'' blocks. the second ones $n_1$ and $n_2$ denote the eigenstate block 1 and 2 initially occupy. The primed and unprimed $i,j,k,l$ are tensor components. 

Since an amorphous solid is isotropic all $3^8$ components of $\chi_{iji'j'}\chi_{klk'l'}$ can be expressed in terms of two independent constants. We chose the longitudinal $\chi_l$ and transverse $\chi_t$ elastic coefficients.
\[\chi_{ijkl}=(\chi_l-2\chi_t)\delta_{ij}\delta_{kl}+\chi_t(\delta_{ik}\delta_{jl}+\delta_{il}\delta_{jk})\]
This holds true for any initial state (to avoid clutter we drop the initial state subscript till eqn(\ref{finaltr})). Let us define $x$,
\[x+2\equiv\frac{\chi_l}{\chi_t}\]
so that
\[\chi_{ijkl}=(x\delta_{ij}\delta_{kl}+\delta_{ik}\delta_{jl}+\delta_{il}\delta_{jk})\chi_t.\]
which simplifies the massive sum of eqn(\ref{trace}) into,
\begin{align}
\!\Lambda_{ijkl}\Lambda_{i'j'k'l'}\chi_{iji'j'}\chi_{klk'l'}=
\!\Lambda_{ijkl}\Lambda_{i'j'k'l'}(x\delta_{ij}\delta_{i'j'}+\delta_{ii'}\delta_{jj'}+\delta_{ij'}\delta_{ji'})\\
\times(x\delta_{kl}\delta_{k'l'}+\delta_{kk'}\delta_{ll'}+\delta_{kl'}\delta_{lk'})\chi_t^2.
\end{align}
The symmetries of $\Lambda_{ijkl}$ simplifies matters a bit further,
\[\Lambda_{ijkl}=\Lambda_{kjil}=\Lambda_{ilkj}=\Lambda_{klij}.\]
Substituting the experimental values \cite{mb} of $\mu_l/\mu_t$ and $\chi_l/\chi_t$ into $\tilde{\Lambda}_{ijkl}$, the entire sum can be evaluated in terms of $x$,
\[\frac{1}{\mu_t^2}\tilde{\Lambda}_{ijkl}\tilde{\Lambda}_{i'j'k'l'}\chi_{0,iji'j'}\chi_{0,klk'l'}\approx122Q^{-2}_{0t}\equiv KQ^{-2}_0\]
a number independent of the relative orientation of the blocks. Thus, eqn(\ref{trace}) becomes
\begin{eqnarray}\label{finaltr}
M=K\frac{v_1v_1}{v_{12}^2}\sum_{n_1n_2}\int\!\!\int\!\! Q_{0,n_1}^{-1}(\omega')Q_{0,n_2}^{-1}(\omega'')d\omega'd\omega''
\end{eqnarray}
where $v_{12}$ is the volume of the sphere with radius equal to the distance between the two blocks. Notice that $M$ is precisely proportional to the square of $\langle Q^{-1}\rangle$. Assuming that the blocks are statistically identical, we can write
\begin{align}\label{t00}
M_0=K\frac{v_av_b}{v_{ab}^2}U_0^2N_b^2\langle Q_{0}^{-1}\rangle^2
\end{align}
Now that we have evaluated $M$ in the eigenbasis of $H_0$ which allowed us to express it in terms of $Q_0$, $N_b$ and $U_0$, we could also repeat the same steps, this time in the eigenbasis of $H=H_0+V$, and express the trace in terms of the above quantities $Q$, $N_b$ and $U$, now modified due to the presence of $V$.
\begin{eqnarray}\label{t0}
M=K\frac{v_1v_1}{v_{12}^2}U^2N_b^2\langle Q^{-1}\rangle^2
\end{eqnarray}
Note that $\langle Q^{-1}\rangle$ is \emph{not} necessarily the averaged absorption of the superblock 1+2 (or even related to it by a simple numerical factor), because the definition of the latter involves the squared matrix elements of the \emph{total} stress tensor of the superblock, $(\hat{T}_{ij}^{(1)}+\hat{T}_{ij}^{(2)})$ and thus contains terms like $\langle n|\hat{T}_{ij}^{(1)}|m\rangle\langle m|\hat{T}_{ij}^{(2)}|n\rangle$ (where $|m\rangle, |n\rangle$ now denote eigenstates of $\hat{H}$); while such terms were originally (in the absence of $\hat{V}$) uncorrelated, it is not obvious that they remain uncorrelated after $\hat{V}$ is taken into account. We shall, however, argue that on average those terms are likely to be small compared to terms of the form $|\langle m|T_{ij}^{(s)}|n\rangle|^2$, because $\hat{V}$ involves \emph{all} tensor components of $\hat{T}^{(1)}$ and $\hat{T}^{(2)}$ while the correlation only involves the \emph{same} component of $\hat{T}^{(1)}$ and $\hat{T}^{(2)}$. If this argument is accepted, we can identify the $\langle Q^{-1}\rangle$ in (\ref{t0}) with the physical inverse absorption of the superblock.

$\mbox{Tr}\hat{V}$ is a scalar, and of course its value is independent of which basis we evaluate it. Equating (\ref{t0}) to (\ref{t00})
\begin{eqnarray}\label{invariance}
U_0^2\langle Q_0^{-1}\rangle^2=U^2\langle Q^{-1}\rangle^2. 
\end{eqnarray}

To relate $\langle Q_0^{-1}\rangle$ to the interacting one $\langle Q^{-1}\rangle$, we must find $(U)/(U_0)$. This will be done using the following argument: Let us square our interacting Hamiltonian of eqn(\ref{2.3.4}) and consider its trace over the same manifold as that defined by the integration limits in (\ref{tr2}).
\begin{align}\label{uncorrelatedT}
\mbox{Tr}(H^2)=\mbox{Tr}(H_0^2)+\mbox{Tr}(H_0V)+\mbox{Tr}(VH_0)+\mbox{Tr}(V^2)
\end{align}
Since pairs of stress tensors are uncorrelated, we can neglect the second and third terms when evaluating the expression in the noninteracting basis. Thus,
\begin{eqnarray}\label{tr1}
\mbox{Tr}(H^2)-\mbox{Tr}(H_0^2)=\mbox{Tr}V^2
\end{eqnarray}
The \textit{lhs} is the change in the variance of energy levels. Since the \textit{rhs} is positive, we can see that the physical effect of the interaction is to spread the energy levels. This equation can simply be written in terms of the density of states and eqn(\ref{t0}),
\begin{eqnarray}\label{dos1}
\int_0^U\omega^2 f(\omega)d\omega-\int_0^{U_0}\omega^2f_0(\omega)d\omega=K\frac{v_av_b}{v_{ab}^2}U_0^2N_b^2\langle Q_{0a}^{-1}\rangle\langle Q_{0b}^{-1}\rangle.
\end{eqnarray}
In general, a macroscopic quantum mechanical system has a density of states roughly given by 
$f(\omega)\propto\omega^n$, where $n$ is a large exponent proportional to the number of constituent particles. However we can afford to be more general and assume
\begin{align}\label{assumef}
f_0(\omega)&=\sum_nc_{0n}\omega^{n_0} \\
f(\omega)&=\sum_nc_{n}\omega^n.
\end{align}
where the dominating terms have power much greater than unity (cf. eqn(\ref{dosintegral})). Note that the experimentally measured temperature dependence of specific heat $C_\nu\sim T$ is consistent with this form for the many-body density of states. In Appendix-B this consistency is shown, and (although not necessary for the present argument) the precise form of $c_n$, as well as an approximate closed form for $f$ is derived. For each set of coefficients $c_n$ and $c_{0n}$ the normalization conditions require that the two body system has $N_b^2$ levels whether they interact or not
\begin{eqnarray}\label{N}
\sum_{n=0}^{p}c_{0n}\frac{U_0^{n_0+1}}{n_0+1}=\sum_{n=0}^{p}c_n\frac{U^{n+1}}{n+1}=N_b^2\nonumber\\
\end{eqnarray}
Since both integrals in eqn(\ref{dos1}) have the form
\[I=\int_0^U\omega^2f(\omega)d\omega=\sum_{n=0}c_n\frac{U^{n+3}}{n+3},\]
if the terms for which $n+3\approx n+1$ dominate the density of states, we can use eqn(\ref{N}) to write
\begin{eqnarray}\label{dosintegral}
I=U^2\sum_{n=0}c_n\frac{U^{n+1}}{n+3}\approx U^2N_b^2.
\end{eqnarray}
Thus, eqn(\ref{dos1}) becomes
\begin{eqnarray}\label{connection}
N_b^2(U^2-U^2_0)=\langle Q_{0a}^{-1}\rangle\langle Q_{0b}^{-1}\rangle N_b^2U_0^2K\frac{v_av_b}{v_{ab}^2}.
\end{eqnarray}
Eqn(\ref{connection}) and eqn(\ref{invariance}) are sufficient to solve for $\langle Q^{-1}\rangle\equiv \langle Q_{a+b}^{-1}\rangle$.
\begin{eqnarray}\label{final}
\langle Q^{-1}\rangle=\left[\frac{1}{\langle Q_{0a}^{-1}\rangle\langle Q_{0b}^{-1}\rangle}+K\frac{v_av_b}{v_{ab}^2}\right]^{-1/2}.
\end{eqnarray}
This equation connects the attenuation coefficient of two non-interacting blocks to that of two interacting blocks. 

\section{Coupling N Generic Blocks: Renormalization Group}
\begin{figure} 
\begin{center} 
\includegraphics[scale=0.5]{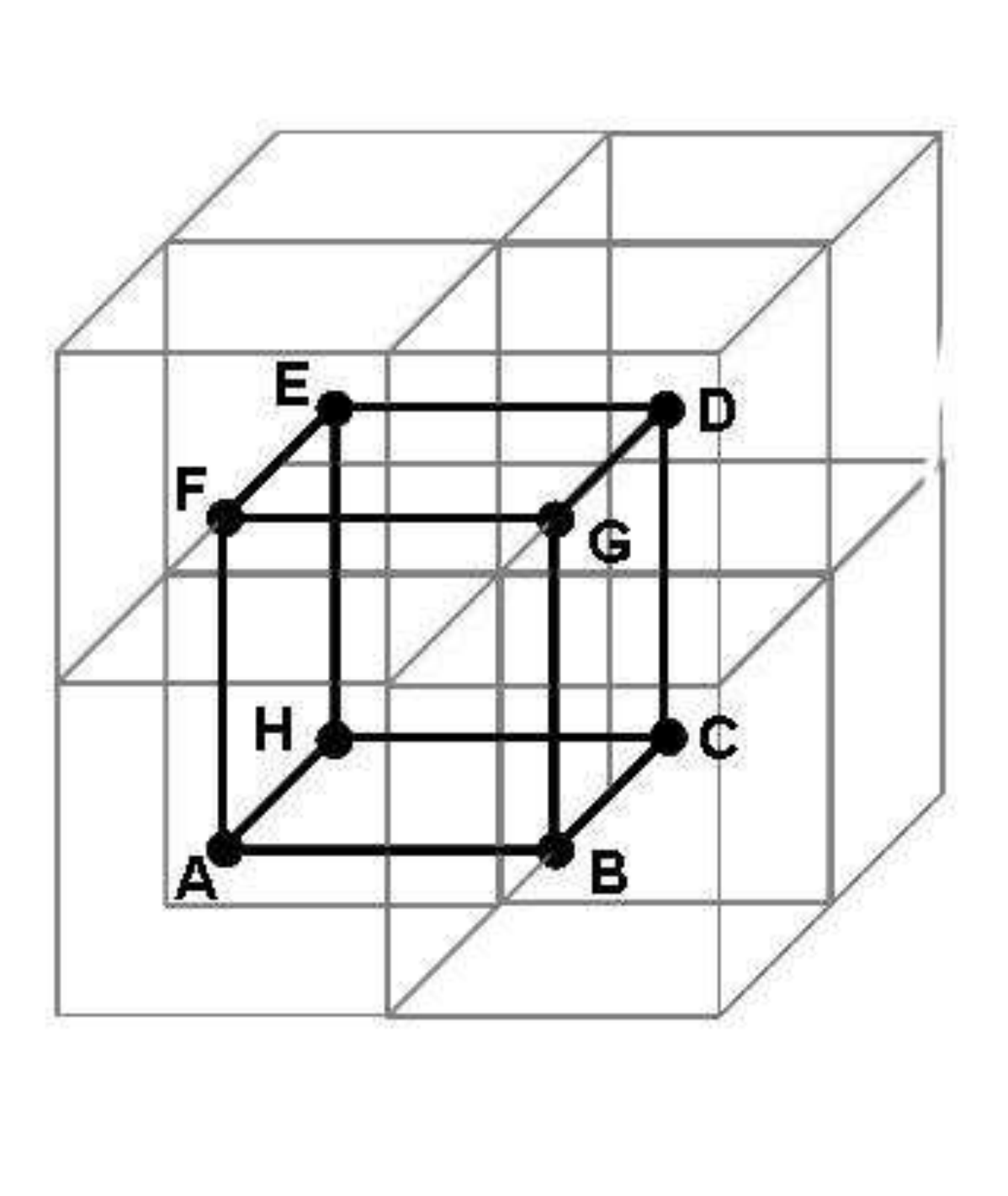} 
\caption{\small \sl A superblock of unit length consisting of 8 blocks.\label{fig:cube}} 
\end{center} 
\end{figure}

We shall use eqn(\ref{final}) to continue adding blocks till we reach experimental length scales. Let start by putting two blocks of side $r_0$ next to each other, so that we have a super-block with dimensions $2r_0\times r_0\times r_0$ (Such as A + H in Fig\ref{fig:cube}). For this super-block ($sb_1$).

\[\left(\frac{v_av_b}{v_{ab}^2}\right)_{sb_1}=\frac{1}{16\pi^2/9}\]
and
\[\langle Q_{sb_1}^{-1}\rangle=\left[\frac{1}{\langle Q_{0}^{-1}\rangle^2}+\frac{9K}{16\pi^2}\right]^{-1/2}.\]
Next we combine two $sb_1$'s to obtain a larger super-block ($sb_2$) with dimensions $2r_0\times 2r_0\times r_0$ (Such as AH + BC in Fig\ref{fig:cube}). This time,
\[\left(\frac{v_av_b}{v_{ab}^2}\right)_{sb_2}=\frac{4}{16\pi^2/9}\]
and
\[\langle Q_{sb_2}\rangle=\left[\frac{1}{\langle Q_{0}^{-1}\rangle^2}+\frac{9K}{16\pi^2}+4\frac{9K}{16\pi^2}\right]^{-1/2}.\]
Finally, we extend one more time to obtain a super-block with the same shape as the original ($sb_3$), but with dimensions $2r_0\times2r_0\times2r_0$ (Such as AHBC + FEGD in Fig\ref{fig:cube}). 
\[\left(\frac{v_av_b}{v_{ab}^2}\right)_{sb_3}=\frac{16}{16\pi^2/9}\]
and denoting $Q_{sb_3}^{-1}=Q_{2L}^{-1}$ and $Q_{0}^{-1}=Q_{L}^{-1}$, we relate the attenuation of a cube with sides $2L$ to that of a cube with sides $L$
\begin{eqnarray}\label{selfsimilar}
\langle Q_{2L}^{-1}\rangle&=&\left[\frac{1}{\langle Q_{L}^{-1}\rangle^2}+\frac{9K}{16\pi^2}+4\frac{9K}{16\pi^2}+16\frac{9K}{16\pi^2}\right]^{-1/2}\nonumber\\
&=&\left[\frac{1}{\langle Q_{L}^{-1}\rangle^2}+K_0\right]^{-1/2}
\end{eqnarray}
where $K_0\approx150$. Eqn(\ref{selfsimilar}) is our central result. It has the very attractive feature that the value of $\langle Q^{-1}\rangle$ is very weakly dependent on $v_0=r_0^3$, and $Q_0^{-1}$, since $K\gg1/\langle Q_0^{-1}\rangle$. We now consider the effect of iterating the step which led to (\ref{selfsimilar}), by combining eight cubes of size $2L$ to make one of size $4L$; for convenience we keep the original definition of the low-energy manifold ($E_m<U_0$), though other choices are also possible. Since the only point at which $U_0$ actually enters the result is implicitly in the definition of the ``average'' in $\bar{Q}^{-1}$ the scale invariant nature of the problem implies that all considerations are exactly the same as at the first stage, and we simply recover (\ref{selfsimilar}) with the replacement of $L$ by $2L$ and $2L$ by $4L$. Continuing the iteration up to a spatial scale $R$, we find
\begin{align}\label{3.2.12}
\langle Q^{-1}(R)\rangle=\left[\langle Q_0^{-1}\rangle^2+K_0\log_2(R/r_0)\right]^{-1/2}
\end{align}
where $r_0$ is the linear dimension of the starting block. 

\begin{figure}[!ht]
\begin{center}
\includegraphics[width=3.6in]{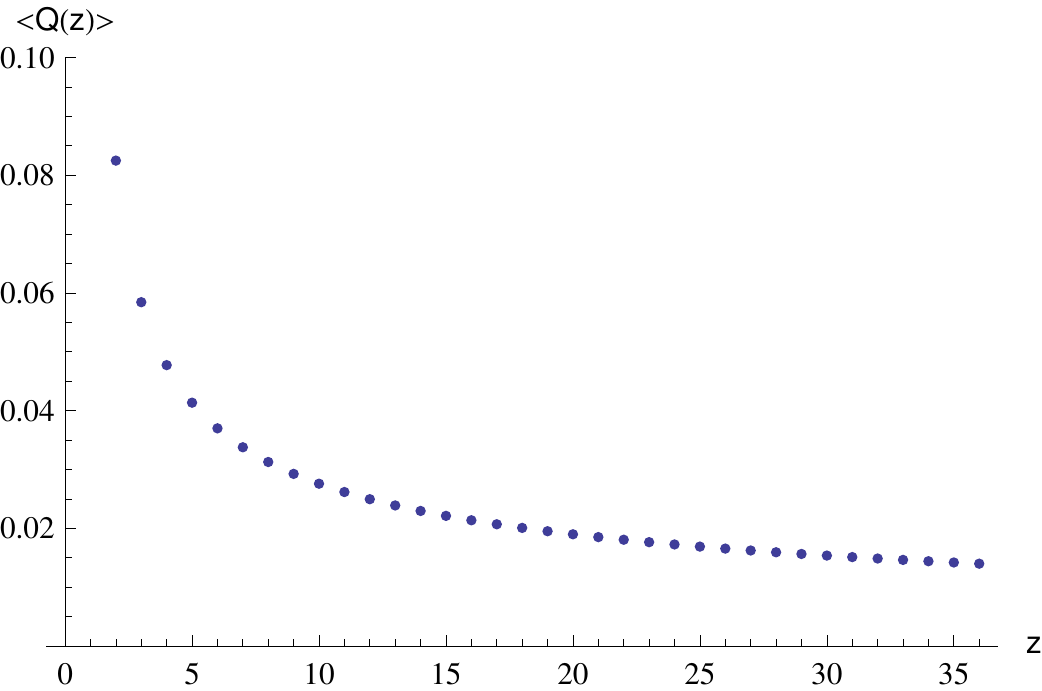}
\end{center}
\caption{
{The (numerical) renormalization of $\langle Q\rangle$}. Each increment in $z$ corresponds to an 8-fold increase in volume. $z=27$ corresponds to $r\sim1$m}\label{fig:vanish}
\label{fig:Figure2}
\end{figure}

Eqn(\ref{3.2.12}) predicts a remarkable counter-intuitive low-temperature effect that would be interesting to test experimentally: As $v\to\infty$, the attenuation vanishes logarithmically (see Fig\ref{fig:vanish}). While we know of no reason why this behavior must be unphysical, in practice we would guess that for any finite ultrasound wavelength $\lambda$, $R$ would be replaced by a quantity of the order $\lambda$. 

It is well known that the temperature dependence of the thermal conductivity of amorphous materials changes at a temperature of the order of $1K$; in fact, most such materials show a pronounced ``plateau'' extending very roughly between 1 and 30K. At 10K the dominant phonons have wavelengths of the order of $50$\AA, and in \cite{leggett91} it is argued that this is just the scale at which we get a crossover from ``Ising'' to ``Heisenberg'' behavior (formally, at smaller scales the approximations used in obtaining the simple $R^{-3}$ form of $\Lambda$, eqn(\ref{2.3.2}) breaks down). Thus, we take the ``starting'' block size, $r_0$, to be $\sim50$\AA (which is still comfortably greater than $a$). Notice that the result (\ref{3.2.12}) depends only logarithmically on $r_0$, and thus the value of $Q^{-1}$ in the experimentally accessible lengthscales will not be particularly sensitive to this choice. Thus, for experimentally realistic values of $R$ we find

\begin{align}\label{result}
\langle Q^{-1} \rangle\sim0.015.
\end{align}

This value is surprisingly small, and more importantly very weakly dependent on the inputs $Q_0$ and $r_0$; however $\langle Q\rangle$ is larger than the experimental values measured in the MHz-GHz range. This is likely due to the contribution to the average value, of the rapid increase in $Q^{-1}$ at higher frequencies, as manifest in the thermal conductivity data around 10K. To obtain the experimentally observed absorption in the MHz-GHz range, we will consider the frequency dependence of $Q^{-1}(\omega)$ in the following chapter.

\chapter{The Frequency Dependence of the Attenuation Coefficient}
\section{Level Shifts}
The reader should be warned that the argument presented in this section is rather heuristic and unorthodox. Suppose that upon turning on the interactions the change in $\chi$ is solely due to the change in the two body density of states $f_0(\omega)$. In other words, we will be ignoring the changes in the matrix elements of $\hat{T}$. The two body density of states $f(\omega)$ after turning on the interactions $V=\Lambda T_1T_2$ can be approximately written as 
\[f(\omega)=\int_0^{U_0}f_0(x)\delta(x-\omega-\Delta(\omega))dx\]
where
\begin{eqnarray}\label{pt}
\Delta(E_n)=\sum_k\frac{V^2_{kn}}{E_n-E_k}
\end{eqnarray}
is the second order correction to the energy level $E_n$. From these two equations,
\begin{align}\label{levelcross}
 \frac{Q^{-1}(\omega)}{Q^{-1}_0(\omega)}\approx\frac{f(\omega)}{f_0(\omega)}\approx\frac{1}{|1+d\Delta(\omega)/d\omega|}
\end{align}
We could write $\Delta$ in terms of $Q$, as described in eqn(\ref{unity})
\begin{align}\label{pertintegral}
\Delta(\omega)=-K_0\int_{-E_{n_10_1}}^\infty\int_{-E_{n_20_2}}^{\infty}\frac{Q_{n_1}(\omega')Q_{n_2}(\omega'')\theta(U_0-|\omega'+\omega''|)}{\omega'+\omega''-\omega}d\omega'd\omega'',
\end{align}
where $E_{n_10_1}+E_{n_20_2}=\omega$. The unit step function $\theta$ imposes the ultraviolet cutoff such that $|E_n-E_k|<U$. In order to take the integral we must know the initial state $|n_1\rangle$ and $|n_2\rangle$ dependence of the absorption. For this, we use the ``random'' form discussed in subsection (2.3.4),
\begin{align}
 Q_n^{-1}(\omega)=\theta(\omega+E_{n})Q^{-1}_0
\end{align}
where the attenuation coefficient of an input block $Q^{-1}_0$ is a constant. This would happen for example, if the level distribution and matrix elements were both uniformly distributed for a starting block. Then the integral in eqn.(\ref{pertintegral}) can be obtained analytically
\begin{eqnarray}\label{delta}
\Delta(\omega)=K_0[-\omega-U-\omega \log(\omega/U_0)]
\end{eqnarray}
Thus, upon turning on the interactions the density of states transforms as
\begin{align}\label{deriv}
 f(\omega)=\frac{f_0(\omega)}{|1+K_0Q^{-2}_0\ln(\omega/U_0)|}
\end{align}
This of course, can be generalized to a ``single shot'' calculation where all blocks within volume $R^3$ contribute. The result is to simply replace the factor $K_0$ by $K_0\log_2(R/L)$:
\begin{align}\label{4.1.5}
Q^{-1}(\omega)&=\frac{Q_0^{-1}}{|K_0Q_0^{-2}\log_2(R/L)\ln(U_0/\omega)-1|}\\
&\approx\mbox{const.}(\ln(U_0/\omega)) \mbox{\,\,\,\,\,\,\,\,if\,\,\,\,} \omega\ll\omega_0
\end{align}

\section{Heuristic Considerations}
While eqn \ref{4.1.5} seems at least qualitatively consistent with the experimental data (see chapter 5), it is not even approximately universal since the constant is inversely proportional to $Q_0^{-1}$, and even given a cutoff at $\omega\sim U_0$ does not satisfy (\ref{3.2.12}). 

There are two obvious reasons form (\ref{4.1.5}) cannot be taken to be literally. First of all, the denominator of (\ref{deriv}) is the absolute value of a negative number. Physically this means that the perturbation is so strong that the levels are crossing. Of course, we know that this cannot happen due to the ``no-level-crossing'' theorem. We will nevertheless assume that when the perturbation is calculated up to infinite order, the overall effect of the higher order terms are small in the level structure, and that the density of states is, at least in the limit $\omega\to0$ qualitatively similar to that given by the first order correction.

The second difficulty is the high-frequency divergence of $\chi$, which indicates that the levels are coming too close. While this too can obviously not be literally true (due to level repulsion) there may be some physical truth in this divergence too, since we know from the thermal conductivity data, that around $T_c$ the phonon mean free path rapidly increases by few orders of magnitude, and the divergence we are seeing in the first order correction might correspond to this rapid rise. We suppose that higher order terms in the perturbation expansion will prevent $Q^{-1}$ from diverging, and therefore place a cut off to the high end by a large value $\tilde{Q}_0^{-1}$ (which need not be equal to the microscopic $Q_0$ defined above, but presumably of the same order of magnitude). Thus, we will consider the following qualitative functional form for the attenuation coefficient
\begin{align}\label{freqdependence}
Q^{-1}(\omega)=\frac{1}{\tilde{Q}_0+A\log(U/\omega)}.
\end{align}

It is possible to support the ansatz (\ref{freqdependence}) further: We know from mean free path measurements, that $Q^{-1}$ is not ``noticeably dependent'' on $\omega$. However if we accept this statement literally, the zero frequency velocity shift as obtained by the Kramers-Kronig formula
\begin{align}\label{kramerskronig}
 \frac{\Delta c}{c}=\Delta\int_0^\infty \frac{Q^{-1}(\omega')d\omega'}{\omega'}
\end{align}
diverges logarithmically. To obtain a finite speed of sound, $Q^{-1}$ must vanish faster with decreasing $\omega$. In fact, it can be shown that the closest form to a constant with non-divergent speed of sound is $Q^{-1}\sim \lim_{epsilon\to0^+}1/\log^{1+\epsilon} (U/\omega)$. 

We shall therefore accept eqn(\ref{freqdependence}) to be the frequency dependence of $Q^{-1}(\omega)$, without necessarily assuming ``its value'' $A$.

\section{The Value of $Q^{-1}(1MHz)$}
Let us now ask what the value of $A$ must be, in order for eqn (\ref{freqdependence}) be consistent with result (\ref{result}). The frequency average can be evaluated analytically,
\[\langle Q\rangle=\frac{1}{U_0}\int_0^{U_0}\frac{d\omega}{1/Q_0+A\log (U_0/\omega)}=\frac{-e^{1/(AQ_0)}}{A}\mbox{Ei}(-1/(AQ_0)).\]
Here, Ei$(x)$ is a special function, defined as
\[\mbox{Ei}(x)=\int_{-\infty}^x\frac{e^t}{t}dt\]
which has the asymptotic form
\[\lim_{x\to0^+}\frac{e^{2\mbox{Ei}(-x)}}{x^2}=e^{2\gamma}\]
where $e^{2\gamma}\approx3.172\ldots$ is the Euler-Mascheroni constant. Thus,
\begin{eqnarray}\label{robust}
\langle Q\rangle=\frac{-e^{1/(AQ_0)}}{2A}\log \left(\frac{e^{2\gamma}}{A^2Q_0^2}\right),
\end{eqnarray}
from which one can obtain the value $A\sim350$ that produces $\langle Q\rangle=0.015$. Now that we know $A$, we may use eqn(\ref{freqdependence}) to find the value of $Q$ in the universal regime. The experimental probing frequencies $\omega$ are typically of the order of MHz, therefore we find
\[Q(\omega=1\mbox{MHz})=2.7\times10^{-4}.\]
which is precisely the ``typical'' experimental value. The dimensionless inverse mean free path $\lambda/l$ is then
\[\lambda/l\approx1/200\]
We emphasize that our central result, namely eqn({\ref{selfsimilar}}), concerns the value of $Q$, which is averaged over $\omega$ and $m$ symmetrically. Therefore if our assumptions on $\omega$ and $m$ were interchanged (or if an assumption pair suitably ``in between'' was used), one would still get similar numbers.

Let us relax some of our assumptions and see how sensitive the value of $Q$ is. For example, if dominating power in the density of states was not a large number, but an arbitrary one, it is not difficult to see that this introduces an extra factor between $1/3$ and $1$ in eqn(\ref{dosintegral}), which in turn alters the value of $\langle Q\rangle$ by a factor of about 1.7, very much within experimental variability per material. Note also that eqn(\ref{robust}) is very robust to fluctuations in $Q_0$. Namely, 
\[\left.\frac{\partial\langle Q\rangle}{dQ_0}\right|_{Q_0\approx1}\ll1.\]
Since the equation (\ref{freqdependence}) was obtained by heuristically, it is also interesting to see how sensitive the value of $Q^{-1}$ is to the precise details of the functional form. For example, consider more general forms that are ``near-constant'', and repeat the above procedure to find $A(s)$.

\begin{align}\label{generalform}
Q^{-1}(\omega)=\frac{1}{[Q_0^{-s}+A(s)\log(U/\omega)]^s}
\end{align}

The dependence of the value of $Q$ to the exponent $s$ and microscopic value $Q_0$ (cf. eqn(\ref{generalform})) is displayed in table-1, which suggests that the universality is more general than that required by the precise form (\ref{freqdependence}).

\begin{table}[h]
\begin{center}
\begin{tabular}{c|c|c|c|c}
  & $s=0.5$ & $s=0.7$ & $s=0.9$\\ \hline
 $Q_0=0.1$ & 0.0024 & 0.0011 & 0.0006\\ \hline
 $Q_0=1$ & 0.0024 & 0.0010 & 0.0004\\ \hline
 $Q_0=10$ & 0.0024 & 0.0010 & 0.0003\\ \hline
\end{tabular}
\caption{The dependence of the average value of $Q^{-1}(1MHz)$ to the details of its functional form}
\end{center}
\end{table}

\chapter{Predictions}
In the previous two chapters we deduced the ultrasonic absorption at zero temperature from generic assumptions and arguments. Unfortunately, to the best of our knowledge no experiment has directly measured $Q^{-1}$ in the regime $T\ll\omega$. Therefore we will compare our results with indirect measurements from which $Q^{-1}$ is deduced. Two such quantities are the thermal conductivity $K$ and the ultrasound velocity shift $\Delta c/c$. As we have done in chapter 4, we will assume that the renormalized $\chi_m(\omega)$ is 
roughly independent of initial state $|m\rangle$ (apart from $\theta(\omega+E_m)$) and hope that a more realistic $m$ dependence will not qualitatively alter the results.
\section{Temperature Dependence of Ultrasonic Velocity Shift}
In the TTLS model, the absorption coefficient is calculated by averaging over that of many TTLS. As discussed in the introduction, in the ``resonance regime'' this is given by
\begin{align}\label{tlsmfp}
 Q^{-1}= Q^{-1}_{hf}\mbox{tanh}\left(\frac{\omega}{2T}\right)
\end{align}
where the coefficient $Q_{hf}^{-1}$ is predicted to be independent of frequency. The resonance behavior of a generic amorphous block is rather similar to that of a TTLS; since $T\ll\omega$ the unexcited distribution of occupied levels are well separated from the excited distribution. Let us therefore designate $p_1$ and $p_2$ as the probability that the amorphous block occupies a level in the former and latter groups respectively. In thermal equilibrium we have, as usual
\begin{align}
p_1+p_2=1\\
\frac{p_2}{p_1}\approx e^{\omega/T}.
\end{align}
The transitions are due to absorption of a phonon and stiumlated emission,
\[\frac{dp_1}{dt}=-W_{12}p_1+W_{21}p_2\]
where the transition rates $W_{12}$ and $W_{21}$ are given by Fermi's Golden Rule. From these equations it is trivial to calculate the phonon lifetime,
\[\tau_{ph}^{-1}=(1-e^{-\omega/T})\sum_{m}\pi\omega Q_{m}^{-1}\frac{e^{-E_m/T}}{Z(T)}\]
Thus, taking into account the assumption regarding initial state inpdependence $Q_m^{-1}$ can be calculated from the phonon mean free path
\[Q^{-1}(\omega,T)=Q^{-1}_{hf}(1-e^{-\omega/T})\]
where now, 
\[Q^{-1}_{hf}\propto\frac{1}{\ln (U_0/\omega)}.\]
Note that for $\omega\gg T$, $Q^{-1}$ is only logarithmically dependent on frequency, and independent of temperature, consistent with experiment \cite{golding76}, and qualitatively similar to the TTLS prediction (\ref{tlsmfp}).
We can find the velocity shift from the Kramers Kronig relation;
\begin{align}\label{velshift}
 \frac{\Delta c}{c}=-\Delta\frac{\chi_0}{\rho c^2}=-\Delta\int_0^{\infty}\frac{d\omega}{\omega}Q^{-1}(\omega,T)
\end{align}
If we insert (\ref{tlsmfp}) into (\ref{velshift}) we obtain for $T\gg\omega$
\begin{align}\label{tlsvelshift}
\frac{\Delta c}{c}=Q_{hf}^{-1}\ln\left(\frac{T}{T_0}\right)
\end{align}
Note that while this form works quite satisfactorily for low temperatures, it significantly departs from experimental data for higher temperatures \cite{golding76b}. The standard way this difficulty is resolved is by introducing an additional fit function to the TTLS density of states $P(\omega)=\bar{P}$, so that 
\[P(\omega)=\bar{P}(1+a\omega^2)\]
where $a$ is a free parameter. The present model does a better job without such additional fit functions. If we treat $Q^{-1}_{hf}$ as a constant the integral (\ref{velshift}) can be taken analytically. 
\begin{figure}[!ht]
\includegraphics[width=3.1in]{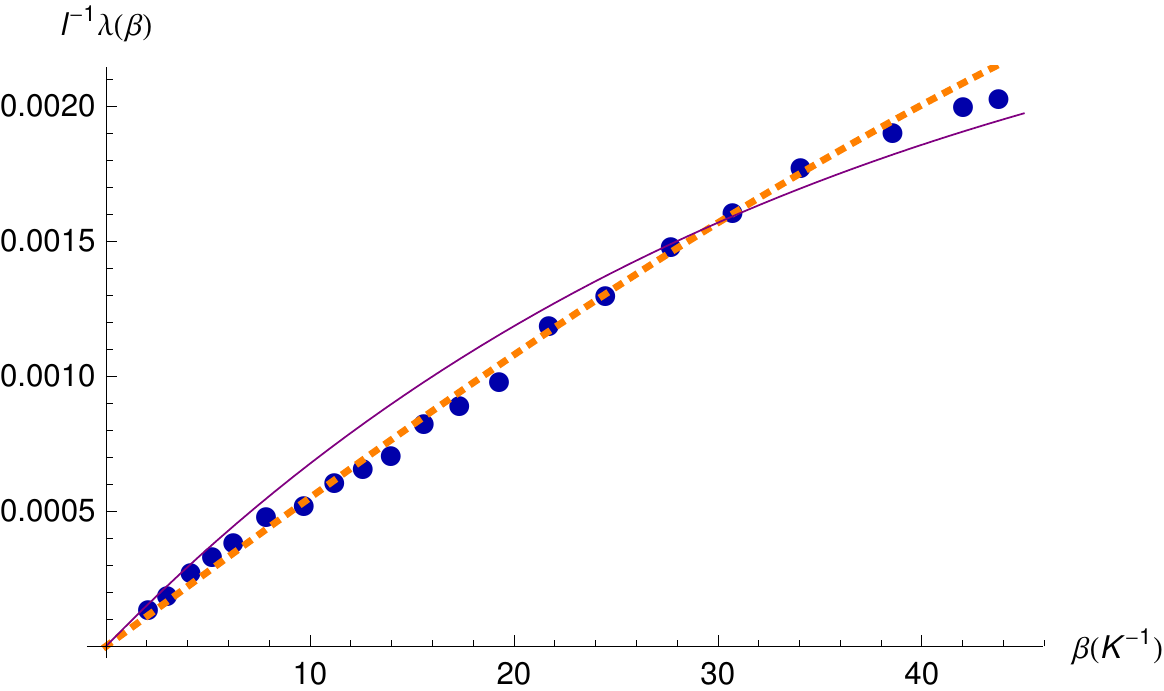}
\includegraphics[width=3.1in]{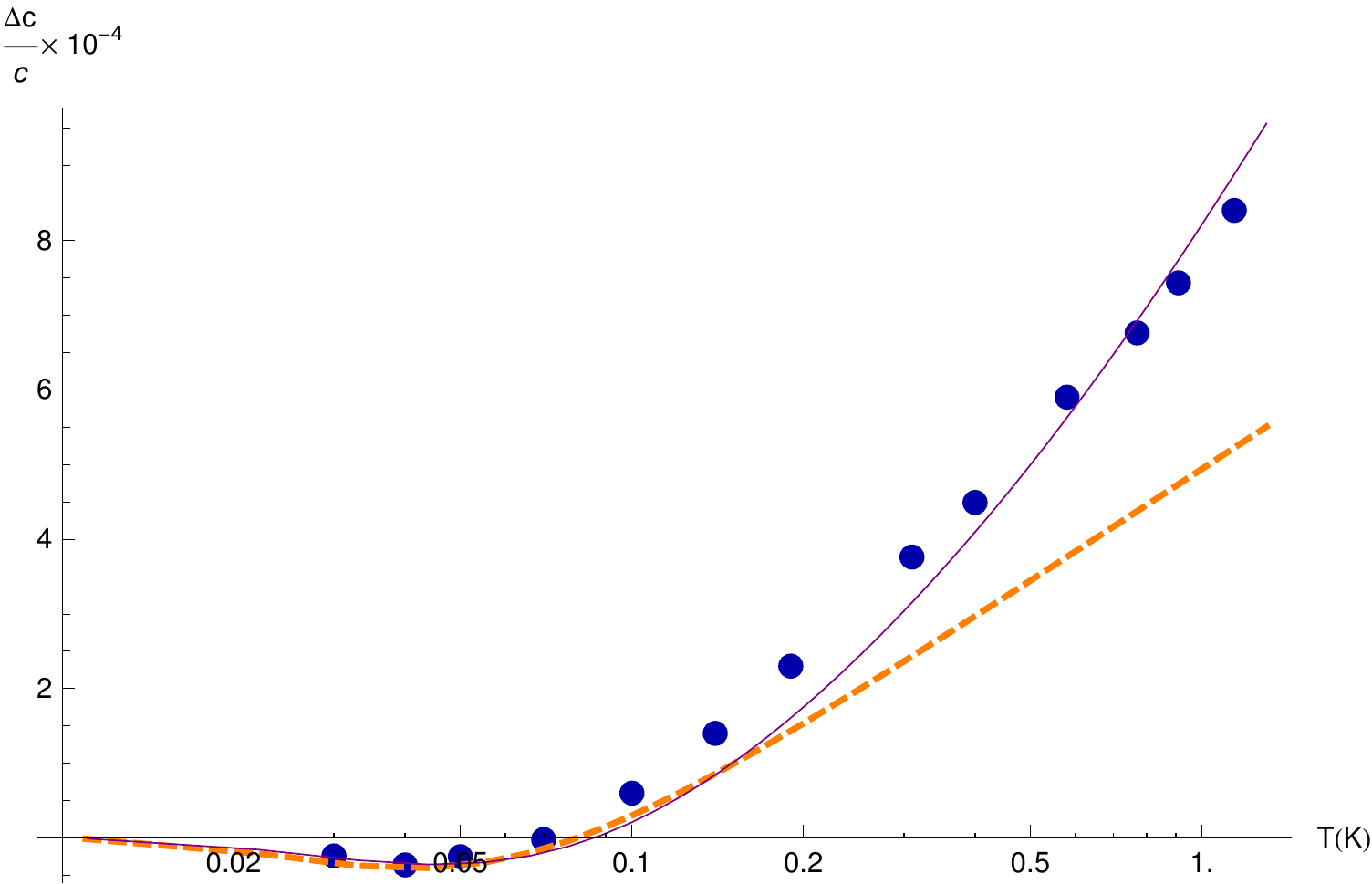}
\caption{
{\bf Normalized inverse mean free path (left) and velocity shift (right).} The present theory (solid) is compared against TTLS without the $\omega^2$ term (dashed) and experiment \cite{golding76,golding76b} (dots). While the functional forms predicted by both models are qualitatively similar at low temperatures, the TTLS model must use an additional fitting function $n(\omega)=n_0(1+a\omega^2)$ for the density of states to resolve the discrepancy in fitting $\Delta c/c$ data (right).}
\label{fig:Figure1}
\end{figure}
\begin{align}
\Delta c/c= \Delta (Q_{hf}^{-1}/2)[-e^{\omega/T}\mbox{Ei}(-\omega/T)-e^{-\omega/T}\mbox{Ei}(\omega/T)]
\end{align}
which is precisely the same result as (\ref{tlsvelshift}) in the $\omega\ll T$ limit. In the presence of $[\ln(U_0/\omega)]^{-1}$, the integration must be done numerically, and is shown in Fig\ref{fig:Figure1} and compared with the TTLS result (with unmodified density of states) as well as experimental data \cite{golding76,golding76b}.

\begin{figure}[!ht]
\includegraphics[width=3.1in]{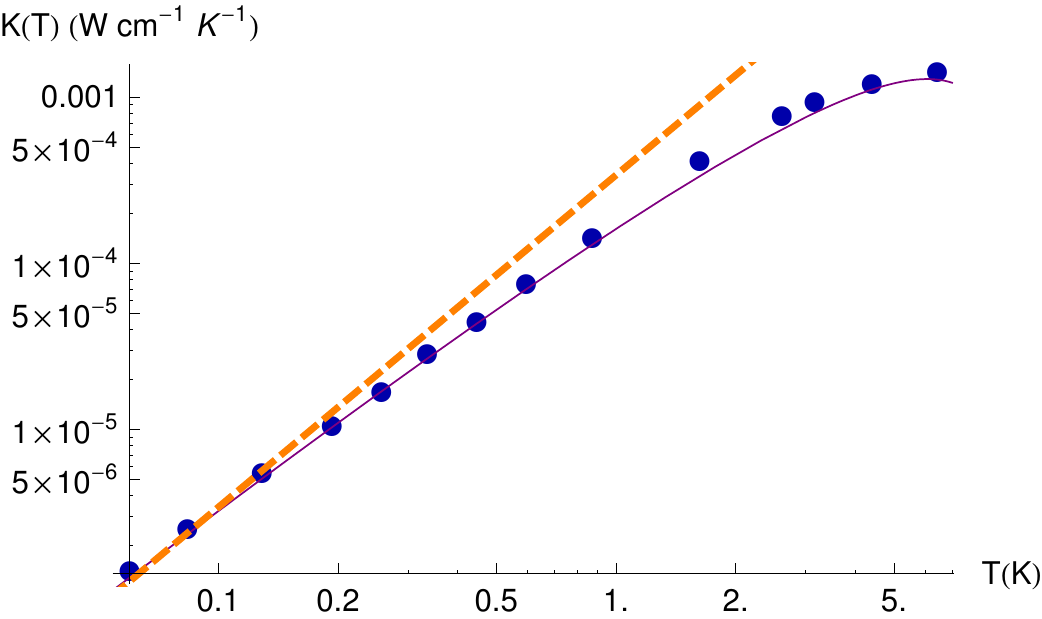}
\caption{
{\bf Temperature Dependence of Thermal Conductivity}. The present theory (solid) $K\sim T^2\ln U_0/T$ is compared against TTLS prediction $K\sim T^2$ and experiment \cite{stephens73} (dots) below the ``plateau''.}
\label{fig:Figure2}
\end{figure}
\section{Temperature Dependence of Thermal Conductivity}
As already mentioned in the introduction, despite the anharmonic degrees of freedom, phonons are the main heat carriers in disordered solids. Thus, using the kinetic formula (\ref{K}),
\begin{align}
K(T)\approx c_\nu^{ph}(T)\bar{c}(T)\bar{l}_{ph}\\
=\mbox{const.} T^2Q(\omega,T)
\end{align}
where $c_\nu^{ph}(T)=\mbox{const.} T^2$ is the phononic specific heat and $\bar{c}(T)$ and $\bar{l}_{ph}$ are the speed of sound and phonon mean free path averaged over frequency. Since at temperature $T$, the frequency distribution function is sharply peaked at $\omega\sim4K$ we can evaluate the frequency averages (``dominant phonon approximation''),
\[K(T)=\mbox{const.}T^2\ln(U_0/T),\]
which is similar to TTLS prediction of $K=\mbox{const.}T^2$, but fits the experimental data better (see Fig\ref{fig:Figure2}).
\newpage 
\newpage\pagebreak
\include{1-introduction}
\include{2-related}
\include{3-model}
\include{4-predictions}

\chapter{Conclusions}

Our main goal in this work has been motivated by providing an explanation to the universal acoustic response of disordered solids at low temperatures, which for us entails a description that does not use ad-hoc fit functions and parameters or postulate other (unobservable) universal quantities or entities. 

We believe that we have been successful in doing so, in that we have shown that (a) simply starting from \emph{arbitrary} uncorrelated blocks and coupling them elastically yields a $\langle Q^{-1}\rangle$ factor that is universal, in the sense that its precise value depends sensitively only to measurable quantities $c_t/c_l$ and $\chi_t/\chi_l$, both of which fluctuate only by a factor of 1.2 across different materials. (b) That given the ansatz derived in chapter 4 for the frequency dependence of $Q(\omega)$, the experimentally probed $Q(1MHz-1GHz)$ is close to the measured value. (c) Other quantities related to $Q^{-1}$, namely the temperature dependences of the mean free path, thermal conductivity and velocity shift, are consistent with our generic model. 
\begin{figure}[!ht]
\includegraphics[width=3.1in]{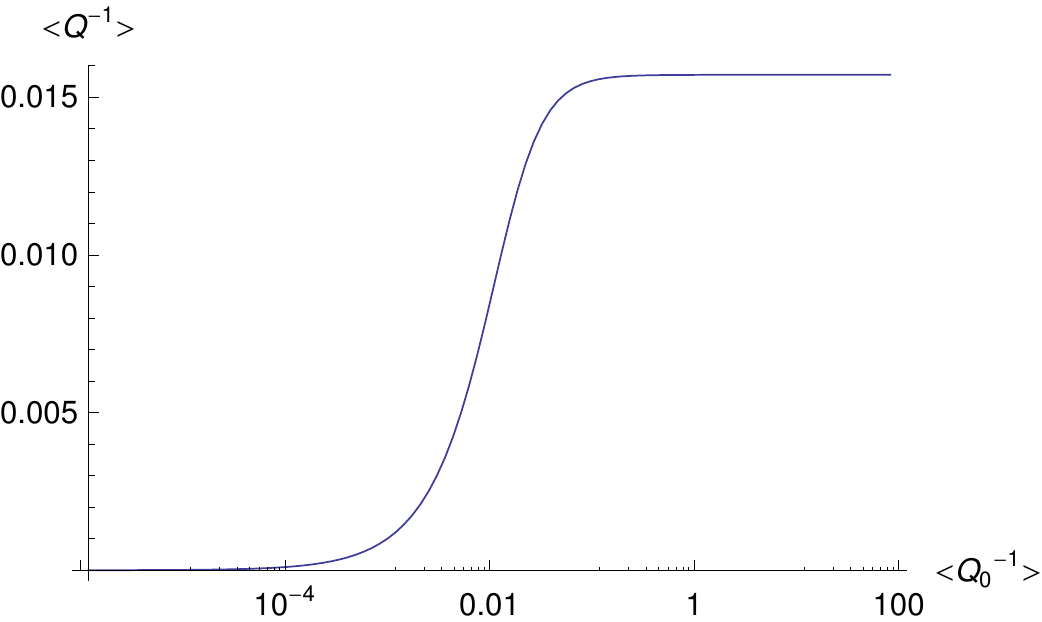}
\caption{
{\bf $\langle Q^{-1}\rangle$ vs $\langle Q_0^{-1}\rangle$ }. Notice that the former can never be larger than $\mathcal{O}(10^{-2})$}
\label{fig:Figure4}
\end{figure}
Finally, we note that the essential ingredient of the present model is the existence of non-phonon degrees of freedom that couples the the strain field linearly. Thus, our arguments should apply equally well to disordered crystals (cf. \cite{stamp09}), with a slightly different angular dependence of the coupling coefficient $\Lambda_{ijkl}$. That being said, our conclusion does not imply that \emph{all} disordered solids must have the same $Q^{-1}$: If the attenuation coefficient of the ``microscopic blocks'' are significantly smaller than the canonical value, then so will be the macroscopic one.

However, a $Q^{-1}$ appreciably larger than the canonical value would tell against our hypothesis (see Fig\ref{fig:Figure4}).
\newpage
\appendix
\chapter{Derivation of the Phonon-Induced Stress-Stress Coupling}
For the sake of completeness we give the derivation of the phonon induced stress-stress coupling defined in section 2 (also, cf. \cite{esquinazi-rev} and \cite{joffrin75}). Consider two generic ``defect modes'', with (non-phonon) stress tensors $\hat{T}_1$ and $\hat{T}_2$ residing at points $\vec{r}_1$ and $\vec{r}_2$, such that they are coupled to the phonon field at $e_{ij}(\vec{r}_1)$ and $e_{ij}(\vec{r}_2)$ linearly. The full Hamiltonian, includes three terms: The defects $H_d$, the phonons $H_{el}$, and the coupling $H_c$
\begin{eqnarray}
H&=&\hat{H}_d+\hat{H}_{el}+\hat{H}_c\nonumber\\
&=&\hat{H}_{d}(\hat{T}_1)+\hat{H}_{d}(\hat{T}_2)+\sum_{q\mu}\left(\frac{|\hat{p}_{q\mu}|^2}{2m}+\frac{m\omega_{q\mu}^2}{2}|\hat{u}_{{q\mu}}|^2\right)\nonumber\\&+&\sum_{\alpha\beta}e_{\alpha\beta}(\vec{r}_1)\hat{T}_{1\alpha\beta}+e_{\alpha\beta}(\vec{r}_2)\hat{T}_{2,\alpha\beta}
\end{eqnarray}
Where $H_d$ corresponds to the internal energy of the defects, $u_{i\alpha}(\vec{r})$ is the $\alpha$ vector-component of the displacement of the atom at position $\vec{r}$, of defect $i$, the subscript $\mu$ indicates the phonon branch and $\hat{u}_{{q\mu}}$ is the Fourier component of the phonon mode with momentum $q$. Namely, if $\vec{e}$ is the unit vector along the direction of $\vec{q}$,
\[u_{1\alpha}=\frac{1}{\sqrt{N}}\sum_{q\mu}U_{q\mu}e_{q\mu\beta}e^{i\vec{q}.\vec{x_1}}.\]
In the formal treatment of phonons in a crystal, $N$ stands for the number of atoms in a unit cell. While defining such a cell (or for that matter defining phonons) is rather questionable for a glass, we may think of $N$ as an (arbitrary) number such that the above equation is meaningful for phonons with wavelength satisfying $N a\gg\lambda$. Let us write $H_c$ explicitly from the definition of strain;
\begin{eqnarray}
 \hat{H}_c=\frac{1}{\sqrt{N}}\sum_{q\mu}u_{q\mu}iq_\beta\epsilon_{q\mu,\alpha}(e^{i\vec{q}.\vec{r}_1})
\end{eqnarray}
The effect of $H_c$ is to shift the equilibrium positions of atoms such that phonon modes can be defined with respect to this equilibrium.
We shall define a new phonon coordinate for each $q$ and $\mu$,
\[\tilde{u}_{q\mu}=u_{q\mu}-u_{0q\mu}\]
in terms of this new equilibrium coordinate $u_{0q\mu}$, which can be found from setting
\[\left(\frac{\partial H}{\partial u_{q\mu}}\right)=0.\]
That is
\[\sum_{q\mu}m\omega_{q\mu}u_{0q\mu}=\frac{-1}{\sqrt{N}}\sum_{q\mu}iq_\beta\epsilon_{q\mu,\alpha}(T_{1,\alpha\beta}e^{i\vec{q}.\vec{x}_1}+T_{2,\alpha\beta}e^{i\vec{q}.\vec{x}_2})\]
Thus,
\[\hat{u}_{0,q\mu}=\frac{-iq_\beta\epsilon_{q\mu,\alpha}}{m\omega_{q\mu}^2\sqrt{N}}(T_{1,\alpha\beta}e^{i\vec{q}.\vec{x}_1}+T_{2,\alpha\beta}e^{i\vec{q}.\vec{x}_2})\]
Then the full hamiltonian can be written in terms of this coordinate,
\[H=H_d+\sum_{q\mu}\left(\frac{p_{q\mu}^2}{2m}+\frac{m\omega_{q\mu}^2}{2}|\tilde{u}_{qm}|^2-\frac{m\omega_{q\mu}^2}{2}|\hat{u}_{0qm}|^2\right)\]
Note that the first two terms in the paranthesis is formally identical to the phonon Hamiltonian, whereas the third, $V_c$ explicitly depends on relative defect positions. That is,
\[V_c=-\sum_{q\mu}\frac{q_\beta q_\delta\epsilon_{q\mu,\alpha}\epsilon_{q\mu,\gamma}}{2m\omega_{q\mu}^2N}(T_{1,\alpha\beta}e^{i\vec{q}.\vec{x}_1}+T_{2,\alpha\beta}e^{i\vec{q}.\vec{x}_1})(T_{1,\gamma\delta}e^{i\vec{q}.\vec{x}_1}+T_{2,\gamma\delta}e^{i\vec{q}.\vec{x}_1})\]
Let us substitute $\omega_{q\mu}=qc_m$, keeping in mind not to sum momentums with $q>2\pi/a$. Separating the position dependent and constant terms $V_c=\mbox{const.}+V(\vec{R})$ with $\vec{R}=\vec{r}_1-\vec{r}_2$, we find
\[V_c=\mbox{const.}-\sum_{q\mu}\frac{q_\beta q_\delta}{q^2}\frac{\epsilon_{q\mu,\alpha}\epsilon_{q\mu,\gamma}}{mNc_\mu^2}\hat{T}_{1,\alpha\beta}\hat{T}_{2,\gamma\delta}\cos(\vec{q}.\vec{R})\]
Of course, constant part is rather uninteresting ($H_d$ could as well be redefined to include these constants, so that the defects are ``dressed'' with phonons). It is the position dependent term that mixes the energy levels of the blocks, and correlates their stress matrices.

For a longitudinal mode $\mu=l$, 
\[e_{ql,\alpha}=\frac{q_\alpha}{|q|}\]
whereas for transverse modes $\mu=t_1,t_2$,
\[e_{qt_1,\alpha}q_\alpha=e_{qt_2,\alpha}q_\alpha=0\]
and
\[\sum_{\mu=t_1t_2}e_{q\mu,\alpha}e_{q\mu,\beta}=\delta_{\alpha\beta}-\frac{q_\alpha q_\beta}{|q|^2}\]
These can be substituted in $\hat{V}(\vec{R})$ when summing over $\mu$
\begin{eqnarray}
 \hat{V}(\vec{R})=\sum_q\frac{q_\beta q_\delta q_\alpha q_\gamma}{mN q^4}\hat{T}_{1,\alpha\beta}\hat{T}_{2,\gamma\delta}\cos(\vec{q}.\vec{R})(\frac{-1}{c_l^2}+\frac{1}{c_t^2})\\
-\sum_q\frac{q_\beta q_\delta \delta_{\alpha\gamma}}{mN c_t^2q^2}\hat{T}_{1,\alpha\beta}\hat{T}_{2,\gamma\delta}\cos(\vec{q}.\vec{R})
\end{eqnarray}
Letting 
\[\sum_q\to \frac{V}{8\pi^3}\int_0^{2\pi/R}d^3q\]
we may evaluate $V(\vec{R})$. Ommiting the angular dependences
\[q_\alpha q_\beta q_\gamma q_\delta/q^4\]
and
\[q_\alpha q_\delta/q^2\]
it can be seen immediately that
\[\hat{V}=\int_0^{2\pi/R}q^2\hat{T}_{1,\alpha\beta}\hat{T}_{2,\gamma\delta}\propto \frac{\hat{T}_{1,\alpha\beta}\hat{T}_{2,\gamma\delta}}{R^3}.\]
The angular dependence can be taken into account to give Eqn. (\ref{interaction}).

\chapter{Many-Body Density of States of Disordered Solids}
This appendix connects assumption (\ref{dosassumption1}) to experiment, particularly the temperature dependence of specific heat $C_v(T)$. We also find the density of states the many-body density state of amorphous solids in closed form. 

The amorphous specific heat, as found from experiment is,
 \[C_v(\beta)=\frac{a}{\beta}.\]
 The average energy is related to the partition function as
 \begin{eqnarray}\label{U}
 U=-\partial_\beta\log Z
 \end{eqnarray}
 Since the specific heat of glasses are linear in temperature, the $U$ must have the form,
 \[U=aT^2+b=\frac{a}{\beta^2}+b\]
 At zero temperature the block is in its ground state, which we (arbitrarily) chose to be zero. Thus
 \[b=0.\]
Then from eqn(\ref{U}),
\begin{eqnarray}\label{partition}
Z=Ae^{a/\beta}
\end{eqnarray}
Here the constant A is the degeneracy of the ground state. This is because
\[\lim_{\beta\rightarrow\infty} Z=\lim_{\beta\rightarrow\infty}\sum_ne^{-\beta E_n}\]
is equal to number of states for which $E_n$ is exactly equal to zero. We can now find the density of states $f(E)$ from
\[Z_\beta=\int_0^\infty f(E)e^{-\beta E}dE\]
by substituting eqn(\ref{partition}) on the lhs and expand both the partition function and the density of states,
\[Ae^{a/\beta}=A(1+\frac{a}{\beta}+\ldots)=\int_0^\infty [f^{(1)}(E)+f^{(2)}(E)+\ldots]e^{-\beta E}dE.\]
The first term of the left hand side can be obtained if
\[f^{(1)}(E)=A\delta(E),\]
whereas the second can be obtained if
\[f^{(2)}(E)=aA.\]
This way we can get $f(E)$ up-to all orders. The result is,
\[f(E)=A[\delta(E)+S(E)]\]
where
\begin{eqnarray}
S(E)=\sum_{n=1}^\infty\frac{E^{n-1}a^n}{n!(n-1)!}.\label{dos}
\end{eqnarray}
One may ask why there is a delta function in the density of states. This is because, for $\epsilon$ close the level spacing, the integral
\[\int_{-\epsilon}^\epsilon f(E)dE\]
must give $A$, the number of levels at $E=0$, consistent with our choice of $E_0$ and eqn(\ref{partition}).
Let's look at for which value of n is the most prominent in eqn(\ref{dos}). We will assume that $k\gg1$, since these are the typical energies accesible to experiment. 
\[S_k=\log\left[\frac{c^kE^{k-1}}{k!(k-1)!}\right]=k\log c+(k-1)\log E-\log k!-\log(k-1)!\]
Using Stirling's approximation, $\log k!=k\log k-k+1$
\[\log S_k=k(\log c+\log E)-\log E-k\log k+2k-3-k\log(k-1)+\log(k-1).\]
differentiating with respect to $k$ and setting it to zero, we find
\[k(k-1)=cE.\]
Thus, at energy $E$, the term $k\approx\sqrt{cE}\gg1$ dominates the density of states.
\[S_{\sqrt{cE}}=2\sqrt{cE}-2\]
One can find a closed form for $f(E)$ by using an asyptotic form of the modified bessel function of the first kind, $I_\nu(z)$. For $z\gg 1$, we can approximate
\[I_1(z)\approx\frac{e^z}{\sqrt{2\pi z}}\]
since the density of states can be written in terms of this special function
\begin{eqnarray}\label{closeddos}
f(E)=A\left[\frac{\sqrt{c}I_1(2\sqrt{cE})}{\sqrt{E}}+\delta(E)\right]
\end{eqnarray}
we can simplify\
\begin{eqnarray}\label{dossimple}
f(E)\approx Ac^{1/4}\frac{e^{2\sqrt{cE}}}{8^{1/4}\pi^{1/2}}
\end{eqnarray}
This asymptotic form of the modified bessel function of the first kind is well known. It is derived from a saddle point approximation, by writing
\[\sum_k S_k\approx\int S_k dk=\int e^{\mbox{ln}{S_k}}dk\]
and expanding $\mbox{ln}S_k$ at its sharp peak up to second order. 

\backmatter
\bibliographystyle{unsrt}
\bibliography{mybib}
\backmatter
\chapter{Author's Biography}

Dervis Can Vural was born on January 10, 1982, in New Brunswick, New Jersey. He received his B.Sc. degree in Physics from Middle East Technical University in Turkey, in 2002 and his M.S degree in Physics from the University of Illinois at Urbana-Champaign in 2004.

\end{document}